\documentclass[10pt]{amsart}

\usepackage{graphicx}
\usepackage{dcolumn}
\usepackage{bm}

\usepackage[utf8]{inputenc}
\usepackage[T1]{fontenc}
\usepackage{mathptmx}

\usepackage{amsthm}
\usepackage{amssymb}
\usepackage{eurosym}
\usepackage[colorlinks,linkcolor=blue,citecolor=blue,urlcolor=blue]{hyperref}

\newcommand{\beq}{\begin{equation}}
\newcommand{\eeq}{\end{equation}}
\newcommand{\bea}{\begin{eqnarray}}
\newcommand{\eea}{\end{eqnarray}}

\newtheorem{theorem}{Theorem}

\newtheorem{definition}[theorem]{Definition}

\theoremstyle{definition}

\newtheorem{remark}[theorem]{\textbf{Remark}}

\newtheorem*{example}{\textbf{Example}}

\begin{document}



\author[Jos\'{e} M. Amig\'{o}]{Jos\'{e} M. Amig\'{o}}
\address{Centro de Investigaci\'{o}n Operativa, Universidad Miguel Hern\'{a}%
ndez, 03202 Elche, Spain}
\email[Corresponding author]{jm.amigo@umh.es}
\author[Roberto Dale]{Roberto Dale}
\address{Centro de Investigaci\'{o}n Operativa, Universidad Miguel Hern\'{a}%
ndez, 03202 Elche, Spain}
\email{rdale@umh.es}

\author[Piergiulio Tempesta]{Piergiulio Tempesta}
\email{p.tempesta@fis.ucm.es, piergiulio.tempesta@icmat.es}
\address{Departamento de F\'{\i}sica Te\'{o}rica, Facultad de Ciencias F%
\'{\i}sicas, Universidad Complutense de Madrid, 28040 -- Madrid, Spain and Instituto de Ciencias Matem\'aticas, C/ Nicol\'as Cabrera, No 13--15,
28049 Madrid, Spain}

\title[ Permutation group entropy ]{Permutation group entropy: a new route
to complexity for real-valued processes}
\maketitle

\date{\today}

\begin{abstract}
This is a review of group entropy and its application to permutation
complexity. Specifically we revisit a new approach to the notion of
complexity in time serie analysis, based on both permutation entropy and
group entropy. As a result, the permutation entropy rate can be extended
from deterministic dynamics to random processes. More generally, our
approach provides a unified framework to discuss chaotic and random
behaviors.
\end{abstract}

\tableofcontents

\maketitle

\bigskip

\textbf{Group entropy is a general concept that includes the perhaps three
best known entropies: Shannon, R\'{e}nyi and Tsallis. More interesting for
us, group entropy can be adapted to any complexity class of systems, the
latter being defined by the asymptotic growth of states in the systems as
the number of their constituents increases. The result is the notion of }$Z$%
\textbf{-entropy of a class, whose rate converges for all systems in that
class. Thus, the R\'{e}nyi entropy (including the Shannon entropy) is the }$%
Z $\textbf{-entropy for the exponential class, while the Tsallis entropy
belongs to systems in the sub-exponential class. All this comes to mind
because permutation entropy is the Shannon entropy of the probability
distributions obtained by quantizing real-valued deterministic processes
(read one-dimensional dynamical systems) via ranks. As it turns out, when
permutation entropy is extended from deterministic processes to random
processes, the complexity class changes from exponential to
super-exponential (factorial, to be more precise) and, hence, the
permutation entropy rate diverges. To solve this problem, we tailor the }$Z$%
\textbf{-entropy for the factorial complexity class to our setting.}

\bigskip

\section{Introduction}

\label{sec:1}

This paper reviews the concept of group entropy and its application to
permutation complexity in general and the ordinal analysis of time series in
particular. In other words, rather than applying the concept of group
entropy to physical systems or abstract complex systems, mostly borrowed
from the statistical mechanics and the general complexity theory, here the
application will focus on a specific symbolic representation of real-valued,
discrete time deterministic and random processes. Let us elaborate a bit on
those concepts.

\textquotedblleft Entropy\textquotedblright\ stands here for a positive
functional on probability distributions such that it fulfills three
conditions, called the first three Shannon-Khinchin (SK) axioms. In the
literature, such functionals usually go by the name of \textit{generalized
entropies} \cite{Amigo2018}. The fourth axiom is called \textit{strong
additivity} or \textit{separability}. The four SK axioms correspond to the
mathematical hypotheses that Shannon and Khinchin used to prove a uniqueness
theorem (up to a positive factor) for the well-known Shannon entropy or, as
we will call it here, the Boltzmann-Gibbs-Shannon (BGS) entropy \cite%
{Shannon,Khinchin}. But, as it turns out, such a uniqueness might be too
restrictive in some situations, specially regarding the \textit{extensivity}
of interacting statistical systems \cite{Tsallis2009,Hanel2011}; this issue
will be discussed in Section \ref{sec:24}. The shortcoming of the BGS
entropy can also be formulated in the following way: its rate only converges
when the number of states grows exponentially with the number of
constituents, as happens in the conventional thermodynamical models and also
in information theory with the number of ever longer words. Extensivity and
growth rate of states are, thus, related concepts that we will find in
several sections below.

That being the case, researchers have considered other properties to
generalize the SBG entropy and, in particular, to replace the fourth SK
axiom. One of the first proposals was \textit{additivity} \cite{Renyi1960}.
The only additive generalized entropy is the R\'{e}nyi entropy, which is
actually a one-parameter family of entropies that includes the BGS entropy
as a limiting case. A more recent proposal that generalizes additivity,
called \textit{composability}, is due to Tsallis \cite{Tsallis2009}. For
example, the one-parameter family of Tsallis entropies are composable; it
includes the BGS entropy as a limiting case, too. The reader interested in
the axiomatic characterization of entropy is referred to the surveys \cite%
{Csiszar2008,Amigo2015}.

Many other approaches to generalized entropies have been proposed in the
literature. In particular, we wish to mention the strongly pseudo-additive
entropies developed in \cite{IS2014PHYSA} and the one advocated in \cite%
{JK2019PRL}, based on statistical inference theory. Indeed, these two
approaches lead to a class of entropies closely related with those
considered in this article, which are defined independently from a
different, group-theoretical perspective. Another interesting approach is
the one proposed in \cite{Hanel2011}, allowing for a classification of
generalized entropies for complex systems based on a scaling approach. Other
formulations are also commonly used in applicative scenarios; we refer the
reader to the very recent review \cite{TS2022ENC} on the modern theory of
generalized entropies.

The specific approach we shall describe here is based on a systematization
of the property of composability proposed by one of the authors, by using
the algebraic theory of formal groups \cite{Haze,Serre1992}. The result is
the concept first proposed in \cite{PT2011PRE} of \textit{group entropy},
that includes the R\'{e}nyi and Tsallis entropies \cite{PT2016AOP,PT2016PRA}%
. A particular class of group entropies, called $Z$-entropies, will play a
central role below: they are tailored to any given complexity class (as
defined by the growth rate of states with the number of constituents) and,
moreover, they are extensive, which means that they asymptotically scale
linearly with the number of constituents, over the uniform probability
distributions.

Permutation complexity arises naturally when real-valued time series are
quantized by means of ranks (equivalently, permutations or \textquotedblleft
ordinal patterns\textquotedblright ) and, more generally, when real-valued,
discrete time deterministic or random processes are represented using those
symbols. Its main tools are ordinal patterns, allowed and forbidden
patterns, and permutation entropy \cite{A2010BOOK}. In particular, the
growth of allowed patterns with the pattern length will play an important
role in our approach. All these concepts will be explained and illustrated
with simple examples as our discussion progresses. The central issue of
permutation complexity that will be addressed here is the divergence of the
conventional permutation entropy rate when applied to random processes.
Precisely, the extensivity of the group $Z$-entropies will come to the
rescue. The result will be a complexity-based permutation entropy, one for
each permutation complexity class (defined by how the number of allowed
ordinal patterns for a process grow with their length), that we call
permutation group entropy.

This review is structured in three parts. The first Part, which comprises
Sections \ref{sec:21}-\ref{sec:26}, is devoted to both the axiomatic
definition of group entropy (Sections \ref{sec:21} and \ref{sec:22}) and the
relevance and main properties of $Z$-entropies in complexity theory
(Sections \ref{sec:23}-\ref{sec:25}). A brief outlook of group entropy is
given in Section \ref{sec:26}.

The second Part, which comprises Sections \ref{sec:31}-\ref{sec:33}, is
devoted to the conceptual and instrumental toolbox of permutation complexity
(Sections \ref{sec:31} and \ref{sec:32}), and permutation complexity classes
(Section \ref{sec:33}). Therefore, these two Parts provide a useful overview
of group entropy and permutation complexity, respectively.

Group entropy and permutation complexity meet in the third Part and both
lead there to the concept of permutation group entropy (Section \ref{sec:41}%
) and its rate (Section \ref{sec:42}), which nicely extends the concept of
permutation entropy rate beyond deterministic processes into the realm of
random processes. Although most of these materials are based on published
work, we have taken special care to present\ them here in a unified and
comprehensive way. Moreover, to make our presentation self-contained, the
fundamental concepts are introduced from scratch while the technical details
can be found in the references given.

\section{Group entropy}

\label{sec:2}

In the last decades, new information measures have been proposed in the
literature, which have found applications in many different contexts arising
from physics, social and economical sciences, as well as classical and
quantum information theory.

In this section, we shall construct new complexity measures based on the
notion of group entropy. As we mentioned in the Introduction, they represent
a large class of generalized entropies, first introduced and discussed in 
\cite{PT2011PRE}, and further extended in a series of papers (see, e.g., 
\cite{PT2016AOP,PT2016PRA,JPPT2018JPA, JT2019ENT, 
RRT2019PRA, TJ2020SR, GPB2020PHA, PT2020CHA, ADT2021CHA, ADT2022CNS, HPPP2022ENT}). In principle, group
entropy represents a versatile tool, that can be used to classify many known
entropies and to design new ones, useful for specific purposes.

\subsection{Beyond the Shannon-Khinchin formulation: the Composability Axiom}

\label{sec:21}

In order to construct a coherent mathematical approach to generalized
entropies, the set of requirements stated by Shannon and Khinchin to
characterize Boltzmann's entropy represent one of the most interesting ways
to formulate a coherent axiomatic approach to the theory.

Let $\{p_{i}\}_{i=1,\cdots ,W}$, $W\geq 2$, be a discrete probability
distribution, i.e., $0\leq p_{i}\leq 1$ and $\sum_{i=1}^{W}p_{i}=1$; we
denote by $\mathcal{P}$ the set of discrete probability distributions with $%
W $ entries for all $W\geq 2$. Let $S$ be a positive function defined on $%
\mathcal{P}$.

The first three \textit{Shannon-Khinchin} (SK) \textit{axioms} essentially
amount to the following properties \cite{Amigo2018}: \newline
\noindent \textbf{(SK1)} \textit{Continuity}: $S(p)$ is continuous with
respect to all variables $p_{1},\ldots ,p_{W}$. \newline
\noindent \textbf{(SK2)} \textit{Maximality}: $S(p)$ takes its maximum value
over the uniform distribution. \newline
\noindent \textbf{(SK3)} \textit{Expansibility}: adding an event of zero
probability does not affect the value of $S(p)$.

These three requirements, nowadays known as the \textit{Shannon-Khinchin
axioms}\textbf{\textbf{\ }}(SK1)-(SK3), represent fundamental,
non-negotiable properties that an entropy $S(p)$ should satisfy to be
physically meaningful. Functions $S(p)$ that satisfy (SK1)-(SK3) are called 
\textit{generalized entropies}.

Historically, axioms (SK1)-(SK3), together with a fourth axiom called strong
additivity (or separability), were introduced by Shannon and Khinchin to
derive the uniqueness of the \textit{Boltzmann-Gibbs-Shannon} (BGS) \textit{%
entropy} 
\begin{equation}
S_{BGS}(p)=S_{BGS}(p_{1},p_{2},...,p_{W})=-\sum_{i=1}^{W}p_{i}\ln p_{i}
\label{S_BGS}
\end{equation}%
(up to a positive factor amounting to the choice of the logarithm base).
Written for simplicity with random variables $X$ and $Y$, \textit{strong
additivity} reads 
\begin{equation}
S(X,Y)=S(X)+S(Y\left\vert X\right) ,  \label{separability}
\end{equation}%
where $S(X,Y)$ is the entropy of the joint probability distribution $p(x,y)$
and $S(Y\left\vert X\right) $, called the \textit{conditional entropy}\ of $%
Y $ given $X$, is the expected value of the entropies $S(Y\left\vert
X=x\right) $.

R\'{e}nyi replaced strong additivity by the weaker property of \textit{%
additivity},%
\begin{equation}
S(p\times q)=S(p)+S(q),  \label{additivity}
\end{equation}%
where $p,q\in \mathcal{P}$ and $p\times q$ is the product probability
distribution. In terms of \textit{independent} random variables $X$ and $Y$
with probability distributions $p(x)$ and $q(y)$, respectively, Equation (%
\ref{additivity}) reads $S(X,Y)=S(X)+S(Y)$; this equation follows from (\ref%
{separability}) because $S(Y\left\vert X\right) =S(Y)$ when $X$ and $Y$ are
independent \cite{Cover2006}. R\'{e}nyi proved \cite{Renyi1960} that the
only additive generalized entropy (up to a positive factor ) is%
\begin{equation}
R_{\alpha }(p)=\frac{1}{1-\alpha }\ln \bigg(\sum_{i=1}^{W}p_{i}^{\alpha }%
\bigg),  \label{Renyi ent}
\end{equation}%
where $\alpha >0$. It holds $\lim_{\alpha \rightarrow 1}R_{\alpha
}(p)=S_{BGS}(p)=:R_{1}(p)$. The one-parameter family $R_{\alpha }(p)$ is now
called the \textit{R\'{e}nyi entropy}.

In contrast, the main idea underlying the concept of group entropy is to
combine axioms (SK1)-(SK3) with the \textit{composability axiom}. The notion
of composability, introduced in \cite{Tsallis2009}, has been put in
axiomatic form in~\cite{PT2016AOP}, \cite{PT2016PRA}, \cite{TJ2020SR} and
related to formal group theory via the notion of group entropy. We shall
briefly discuss these concepts of composability and group entropy as in~\cite%
{PT2016AOP}, \cite{TJ2020SR} to illustrate the potential relevance of the
group-theoretical machinery\textbf{\ }in the study of composite statistical
systems. We also mention that the relevance of the notion of group entropy
in classical information geometry has been elucidated in \cite{RRT2019PRA}.

\begin{definition}[Composability Axiom]
\label{composab} \label{Def1}We say that a generalized entropy $S$ is 
\textit{composable} if there exists a continuous function of two real
variables $\Phi (x,y)$, called the composition law, such that the following
four conditions are satisfied.

\begin{description}
\item[(C1)] Composability: $S(p\times q)=\Phi (S(p),S(q))$ for any $p,q\in 
\mathcal{P}.$ \label{con:com}

\item[(C2)] Symmetry: $\Phi (x,y)=\Phi (y,x)$. \label{con:symm}

\item[(C3)] Associativity: $\Phi (x,\Phi (y,z))=\Phi (\Phi (x,y),z)$. \label%
{con:asoc}

\item[(C4)] Null-composability: $\Phi (x,0)=x$. \label{con:null}
\end{description}
\end{definition}

Choose $\Phi (x,y)=x+y$ to realize that composability trivially includes
additivity. Furthermore, observe that the mere existence of a function $\Phi
(x,y)$ taking care of the composition process as in condition (C1) is
necessary, but not sufficient to ensure that a given entropy may be suitable
for information-theoretical or thermodynamic purposes; this function must
also satisfy the other requirements above to be admissible. Indeed, in
general the entropy of a system compounded by two subsystems $A$ and $B$,
described by the probability distributions $p$ and $q$ respectively, should
not vary if the labels $A$ and $B$ are exchanged, thus justifying
condition~(C2). In the same vein, condition~(C3) guarantees the
composability of more than two systems in an associative way, this property
being crucial to define a zeroth law. Finally, condition~(C4) is, in our
opinion, also necessary since if we compound two systems $A$ and $B$ and the 
$B$ has zero entropy, then the total entropy must coincide with the entropy
of $A$.

The set of requirements (C1)-(C4) altogether constitutes the \textit{%
composability axiom}. The substitution of additivity by composability in the
four Shannon-Khinchin axioms allows us to generalize the concept of entropy
in a useful and interesting way, as we will see.

\begin{definition}
\label{def:groupentropy} A group entropy is a function $S:\mathcal{P}%
\rightarrow \mathbb{R}^{+}\cup \{0\}$ which satisfies the Shannon-Khinchin
axioms (SK1)-(SK3) and the composability axiom (C1)-(C4).
\end{definition}

Our construction of group entropy is inspired by the notion of formal group
law, that we shall define below.

\subsection{Formal group laws}

\label{sec:22}

In the following construction, all ring considered are associative
commutative unital rings. In particular, given a ring $(A,+,\cdot )$, we
shall denote by $0$ the neutral element of the addition operation $+:A\times
A\rightarrow A$ and by $1$ the neutral element of the multiplication
operation $\cdot :A\times A\rightarrow A$.

\begin{definition}
Let $R$ be a commutative ring with identity, and $R\left\{
x_{1},x_{2},..\right\} $ be the ring of formal power series in the variables 
$x_{1}$, $x_{2}$, ... with coefficients in $R$. A commutative
one-dimensional formal group law over $R$ is a formal power series in two
variables $\Phi \left( x,y\right) \in R\left\{ x,y\right\} $ of the form $%
\Phi \left( x,y\right) =x+y+$ \emph{terms of higher degree}, such that 
\begin{equation*}
i)\qquad \Phi \left( x,0\right) =\Phi \left( 0,x\right) =x
\end{equation*}%
\begin{equation*}
ii)\qquad \Phi \left( \Phi \left( x,y\right) ,z\right) =\Phi \left( x,\Phi
\left( y,z\right) \right) \text{.}
\end{equation*}

When $\Phi \left( x,y\right) =\Phi \left( y,x\right)$, the formal group law
is said to be commutative.
\end{definition}

For any formal group law $\Phi (x,t)$, there exists an \textit{inverse
formal series} $\varphi \left( x\right) $ $\in R\left\{ x\right\} $ such
that $\Phi \left( x,\varphi \left( x\right) \right) =0$. This is the reason
why we talk about \textit{formal group laws}. See \cite{Haze} for a thorough
exposition on formal group theory and~\cite{Serre1992} for a shorter
introduction to the topic.

Some relevant examples are listed below.

\begin{itemize}
\item The \textit{additive} \textit{group law} 
\begin{equation*}
\Phi (x,y)=x+y\ .
\end{equation*}

\item The \textit{multiplicative} \textit{group law} 
\begin{equation*}
\Phi (x,y)=x+y+axy\ .
\end{equation*}

\item The \textit{hyperbolic} \textit{group law} (addition of velocities in
special relativity) 
\begin{equation*}
\Phi (x,y)=\frac{x+y}{1+xy}\ .
\end{equation*}

\item The \textit{Euler group law} 
\begin{equation*}
\Phi (x,y)=(x\sqrt{1-y^{4}}+y\sqrt{1-x^{4}})/(1+x^{2}y^{2})\ .
\end{equation*}%
which appears in the sum of elliptic integrals 
\begin{equation*}
\int_{0}^{x}\frac{dt}{\sqrt{1-t^{4}}}+\int_{0}^{y}\frac{dt}{\sqrt{1-t^{4}}}%
=\int_{0}^{\Phi (x,y)}\frac{dt}{\sqrt{1-t^{4}}}\ .
\end{equation*}
\end{itemize}

In the construction of a group entropy, the composition law is algebraically
a commutative formal group law, i.e., a continuous function of the form 
\begin{equation}
\Phi (x,y)=\sum_{n=0}^{\infty }\sum_{i=0}^{n}a_{i,n-i}x^{i}y^{n-i}
\label{Phi(x,y)}
\end{equation}%
that satisfies conditions (C2)-(C4) of Definition \ref{Def1}). Therefore, $%
a_{i,n-i}=a_{i-n,i}$ for all $n\geq 1$ and $0\leq i\leq n$ by symmetry, and $%
a_{0,0}=0$ and $a_{n,0}=0$ for all $n\geq 2$ by the null-composability. This
is the origin of the connection between entropic measures and formal group
theory (presented for the first time in \cite{PT2011PRE}), as we shall
illustrate in the forthcoming considerations.

\subsection{Group logarithms and Z-entropies}

\label{sec:23}

Our construction relies on the notion of group logarithm associated to every
formal group law. There is a certain freedom concerning the regularity
properties in its definition, depending on the application under
consideration. The standard logarithm is associated to the additive formal
group law.

\begin{definition}
A group logarithm is a strictly increasing and strictly concave function $%
\log _{G}:(0,\infty )\rightarrow \mathbb{R}$, with $\log _{G}(1)=0$
(possibly depending on a set of real parameters), satisfying a functional
equation of the form 
\begin{equation}
\log _{G}(xy)=\chi (\log _{G}(x),\log _{G}(y))  \label{glog}
\end{equation}%
where $\chi (x,y)$ fulfills the requirements $(C2)$-$(C4)$. Equation (\ref%
{glog}) will be called the group law associated with $\log _{G}(\cdot )$.
\end{definition}

From now on, we shall focus on group logarithms of the form 
\begin{equation}
\log _{G}(x)=G(\ln x)\ .  \label{eq:GL}
\end{equation}%
Here $G$ is a strictly increasing function of the form $G(t)=t+O(t^{2})$ for 
$t\rightarrow 0$ (hence, vanishing at $0$) and assuring concavity of $\log
_{G}(x)$; see \cite{PT2016PRA} for a general discussion of sufficient
conditions for the function $G$ to comply with these requirements.

A first, relevant example of a nontrivial group logarithm is given by the so
called $q$-logarithm. We have 
\begin{equation}
G(t)=\frac{\mathrm{e}^{(1-q)t}-1}{1-q},\,\log _{q}(x)=G(\ln x)=\frac{%
x^{1-q}-1}{1-q},\,q>0\,.  \label{qlog}
\end{equation}%
\noindent This logarithm has been largely investigated in connection with
nonextensive statistical mechanics \cite{Tsallis2009}. The formal inverse of
a group logarithm will be called the associated group exponential; it is
defined by 
\begin{equation}
\exp _{G}(x)=e^{G^{-1}(x)}.  \label{Gexp}
\end{equation}%
When $G(t)=t$, we obtain the standard exponential; when, as before $G(t)=%
\frac{e^{(1-q)t}-1}{1-q}$, we recover the $q$-exponential 
\begin{equation*}
e_{q}(x)=\left[ 1+(1-q)x\right] _{+}^{\frac{1}{1-q}}\,,
\end{equation*}%
and so on. Infinitely many other examples of group logarithms and
exponentials are provided, for instance, in \cite{PT2011PRE}.

Generalized logarithms are a key ingredient in the construction of group
entropies, and allow us to realize the composition laws for the entropies in
terms of formal group laws. An important instance of the family of group
entropies is the class of $Z$-entropies defined in~\cite{PT2016PRA}. Their
general form, for $\alpha >0$, is 
\begin{equation}
Z_{G,\alpha }(p):=\frac{1}{1-\alpha }\log _{G}\Big(\sum_{i=1}^{W}p_{i}^{%
\alpha }\Big),  \label{eq:Z}
\end{equation}%
where $\log _{G}$ is a group logarithm \footnote{%
The notion of Z-entropy should not be confused with that of z-entropy,
recently introduced in \cite{Liu2022}.}.

\subsection{A general theorem}

\label{sec:24}

Let us introduce the function $\mathcal{W}=\mathcal{W}(N)$ which describes
asymptotically the number of allowed states of a (statistical, complex, ...)
system as a function of the number of particles (constituents, degrees of
freedom, etc.) $N$. We shall call it the state space growth rate function%
\footnote{%
Since we are interested essentially in the large $N$ limit, we can always
think of $\mathcal{W}(N)$ as an integer number (i.e. we shall identify it
with its integer part).} The set of all possible systems sharing the same
growth rate function $\mathcal{W}=\mathcal{W}(N)$ defines a \textit{%
universality class} of systems.

For instance, very many physical systems are characterized by a growth rate
function of the form $\mathcal{W}(N)=k^{N}$, $k>0$. Other natural choices
are $\mathcal{W}(N)=N^{\alpha }$ and $\mathcal{W}(N)=N!$. Generally
speaking, the universality classes can be organized into three families: the 
\textit{sub-exponential}, the \textit{exponential} or the \textit{%
super-exponential family} depending on whether $\mathcal{W}(N)<e^{N}$, $%
\mathcal{W}(N)=e^{N}$, or $\mathcal{W}(N)>e^{N}$ for large $N$,
respectively. A priori, in each of these three families there are
(infinitely) many classes, although, not necessarily realized in terms of
known complex systems.

In \cite{TJ2020SR} the following \textit{extensivity requirement} was
introduced.

\medskip

\textbf{Extensivity Requirement}. \textit{Given an isolated system in its
most disordered state (the uniform distribution), the amount of its disorder
increases proportionally to the number }$N$\textit{\ of its constituents}.

\medskip

In other words, we shall require that, if $S$ is an information measure of
order/disorder for that system, we must have $S(\frac{1}{\mathcal{W}(N)},...,%
\frac{1}{\mathcal{W}(N)})/N=\,$\textrm{const\thinspace }$>0$. For example, $%
R_{\alpha }(\frac{1}{\mathcal{W}(N)},...,\frac{1}{\mathcal{W}(N)})=\ln 
\mathcal{W}(N)$ for $\alpha >0$, which readily shows that $R_{\alpha }(p)$
(including $S_{BGS}(p)$ for $\alpha =1$) is extensive for the exponential
class $\mathcal{W}(N)=e^{N}$ (or $\mathcal{W}(N)=e^{cN}$, $c>0$, for that
matter). A weaker condition, suitable for macroscopic systems, is the
asymptotic condition 
\begin{equation}
\lim_{N\rightarrow \infty }\frac{S(\frac{1}{\mathcal{W}(N)},...,\frac{1}{%
\mathcal{W}(N)})}{N}=\text{\emph{const}}>0.  \label{ext}
\end{equation}%
We stress that in this paper we are not considering thermodynamics, but a
purely information-theoretical context.

In \cite{TJ2020SR}, it has been shown that for any universality class we can
construct in a purely deductive and axiomatic way an entropic functional
representing a suitable information measure for that class. Indeed, this
functional is a group entropy and satisfies the extensivity requirement. The
main result is the following.

\begin{theorem}
\label{theo1} \label{Thm Z-ent}\label{Thm_main}Let $\mathcal{W}$ be the
state space growth rate function, corresponding to a given universality
class of a statistical system. Then there exists a group entropy which
satisfies the extensivity requirement. This entropy is given by 
\begin{equation}
Z_{\mathcal{W},\alpha }(p)=\lambda \left( \mathcal{W}^{-1}\left( \bigg(%
\sum_{i=1}^{W}p_{i}^{\alpha }\bigg)^{\frac{1}{1-\alpha }}\right) -\mathcal{W}%
^{-1}(1)\right) \ ,  \label{maineq}
\end{equation}%
where $\alpha >0$, $\lambda =\frac{1}{(\mathcal{W}^{-1})^{\prime }(1)}$, $%
p=(p_{1},...,p_{W})$ and it is assumed $(\mathcal{W}^{-1})^{\prime }(1)\neq
0 $.
\end{theorem}

\noindent Entropy \eqref{maineq} belongs to the $Z$-class, meaning that it
can be written in the form \eqref{eq:Z} for 
\begin{equation}
G(t)=\lambda (1-\alpha )\left( \mathcal{W}^{-1}\left( e^{\frac{t}{1-\alpha }%
}\right) -\mathcal{W}^{-1}(1)\right) .  \label{eq:Gt}
\end{equation}%
Alternatively, we say that $Z_{\mathcal{W},\alpha }(p)$ is a $Z$-entropy.
The group law associated to $Z_{\mathcal{W},\alpha }(p)$ is given by 
\begin{eqnarray}
\Phi (x,y) &=&\lambda \left\{ \mathcal{W}^{-1}\Big[\mathcal{W}\left( \frac{x%
}{\lambda }+\mathcal{W}^{-1}(1)\right) \mathcal{W}\left( \frac{y}{\lambda }+%
\mathcal{W}^{-1}(1)\right) \Big]\right.  \notag \\
&&\left. -\mathcal{W}^{-1}(1)\right\} \ .  \label{Phi_W}
\end{eqnarray}

\subsection{Examples}

As we have already mentioned, the \textit{exponential class }is defined by
the growth rate function $\mathcal{W}(x)=e^{cx}$, $c>0$. In this case, 
\begin{equation*}
\mathcal{W}^{-1}(s)=\ln (s/c),\;\mathcal{W}^{-1}(1)=\ln \frac{1}{c},\;(%
\mathcal{W}^{-1})^{\prime }(1)=1
\end{equation*}%
and%
\begin{equation*}
\mathcal{W}^{-1}\left( \bigg(\sum_{i=1}^{W}p_{i}^{\alpha }\bigg)^{\frac{1}{%
1-\alpha }}\right) =\ln \left( \frac{1}{c}\bigg(\sum_{i=1}^{W}p_{i}^{\alpha }%
\bigg)^{\frac{1}{1-\alpha }}\right) ,
\end{equation*}%
thus%
\begin{equation}
Z_{\mathcal{W},\alpha }(p)=\frac{1}{1-\alpha }\ln \bigg(%
\sum_{i=1}^{W}p_{i}^{\alpha }\bigg)=R_{\alpha }(p),  \label{Renyi ent2}
\end{equation}%
i.e., the $Z$-entropy for the exponential class is the R\'{e}nyi entropy,
Equation (\ref{Renyi ent}).

Comparison of Equations (\ref{Renyi ent2}) and (\ref{eq:Z}) shows that $%
Z_{G,\alpha }(p)$ is a \textquotedblleft deformation\textquotedblright\ of
the R\'{e}nyi entropy in the sense that the function $\ln
(\sum_{i=1}^{W}p_{i}^{\alpha })$ is replaced by $\ln
_{G}(\sum_{i=1}^{W}p_{i}^{\alpha })$. Indeed, plugging $\mathcal{W}%
(x)=e^{cx} $ into Equation (\ref{eq:Gt}), one gets $G(t)=t$ so that $\ln
_{G} $ becomes the usual Naperian logarithm (see definition (\ref{eq:GL}))
and, hence, $Z_{G,\alpha }(p)$ becomes $R_{\alpha }(p)$.

The growth rate function $\mathcal{W}(x)=x^{\beta }$, $\beta \in \mathbb{R}$%
, defines the \textit{sub-exponential class}. In this case, 
\begin{equation*}
\mathcal{W}^{-1}(s)=s^{1/\beta },\;\mathcal{W}^{-1}(1)=1,\;(\mathcal{W}%
^{-1})^{\prime }(1)=\frac{1}{\beta }
\end{equation*}%
and%
\begin{equation*}
\mathcal{W}^{-1}\left( \bigg(\sum_{i=1}^{W}p_{i}^{\alpha }\bigg)^{\frac{1}{%
1-\alpha }}\right) =\bigg(\sum_{i=1}^{W}p_{i}^{\alpha }\bigg)^{\frac{1}{%
\beta (1-\alpha )}},
\end{equation*}%
thus 
\begin{equation}
Z_{\mathcal{W},\alpha }(p)=\beta \left( \bigg(\sum_{i=1}^{W}p_{i}^{\alpha }%
\bigg)^{\frac{1}{\beta (1-\alpha )}}-1\right) .  \label{2parTE}
\end{equation}

Entropy \eqref{2parTE} was derived in \cite{JT2019ENT} (see Equation (19))
and it can be regarded as a nontrivial two-parametric generalization of
Tsallis entropy. Indeed, it reduces to \textit{Tsallis entropy} \cite%
{Tsallis2009} 
\begin{equation}
T_{\alpha }(p)=\frac{1}{1-\alpha }\left( \sum_{i=1}^{W}p_{i}^{\alpha
}-1\right) \   \label{TE}
\end{equation}%
for $\beta =\frac{1}{1-\alpha }$. It holds $\lim_{\alpha \rightarrow
1}T_{\alpha }(p)=S_{BGS}(p)=:T_{1}(p)$. However, the relationship $\beta =%
\frac{1}{1-\alpha }$ excludes $T_{1}(p)=S_{BGS}(p)$ from the sub-exponential
class.

Among the many remarkable properties of entropy \eqref{TE}, we remind the
following result, proved in \cite{ET2017JSTAT}: Under mild technical
hypotheses, Tsallis entropy is the most general trace-form composable
entropy. An entropy $S(p)=S(p_{1},...,p_{W})$ is said to be \textit{trace
form} if $S(p)=\sum_{i=1}^{W}g(p_{i})$, i.e., it is the sum of univariate
terms, such as the BGS entropy (\ref{S_BGS}) but unlike the R\'{e}nyi
entropy (\ref{Renyi ent}). Equivalently, $T_{\alpha }(p)$ is the most
general trace-form group entropy.

The \textit{super-exponential class}, the most relevant for the forthcoming
analysis, will be considered extensively from Section \ref{sec:32}. The
prototypical entropy of this class is the entropy defined by one of us in 
\cite{JPPT2018JPA} 
\begin{equation}
Z_{\alpha }(p)=\exp \left[ \mathcal{L}\left( \frac{1}{c}R_{\alpha
}(p)\right) \right] -1\ ,  \label{JTTP}
\end{equation}%
which is extensive for $\mathcal{W}(x)=x^{cx}=e^{cx\ln x}$, $c>0$. Here $%
\mathcal{L}(x)$ is the \textit{principal} branch of the real $W$-\textit{%
Lambert function} (i.e., the unique solution of $W(x)e^{W(x)}=x$ for $x\geq
0 $). The basic properties of $\mathcal{L}(x)$ include the following \cite%
{Olver2010}: (i) $\mathcal{L}(x)$ is strictly increasing and $\cap $-convex;
(ii) $\mathcal{L}(-e^{-1})=-1$ and $\mathcal{L}(0)=0$; (iii) $\mathcal{L}%
(x)>0$ for $x>0$; (iv) $\mathcal{L}(x)>1$ for $x>e$; and (v) $\mathcal{L}%
(x)\rightarrow \infty $ as $x\rightarrow \infty $. Moreover, $\mathcal{L}(x)$
satisfies the identity 
\begin{equation}
\mathcal{L}(x\ln x)=\ln x  \label{identity}
\end{equation}%
for $x\geq e^{-1}$.

Another interesting example is the $Z_{a,b}$-entropy, due to its generality.
It is defined to be the function 
\begin{equation}
Z_{a,b}(p_{1},\ldots ,p_{W})=\frac{\left( \sum_{i=1}^{W}p_{i}^{\alpha
}\right) ^{a}-\left( \sum_{j=1}^{W}p_{j}^{\alpha }\right) ^{b}}{%
(a-b)(1-\alpha )}.
\end{equation}%
Here $0<\alpha <1$, and $a>0$, $b\in \mathbb{R}$ or $a\in \mathbb{R}$, $b>0$%
, with $a\neq b$. This entropy was introduced in \cite{PT2016PRA} and is
related to the Abel formal group law \cite{Haze}. This entropy generalizes
the well-known Sharma-Mittal entropy \cite{Sharma1975,Mittal1975}, obtained
for $b=0$, which in turn generalizes R\'{e}nyi's entropy. Besides, when $%
a=b=k$, one obtains a new, composable version of Kaniadakis entropy \cite%
{Kaniadakis2002}. Tsallis entropy is also recovered for $b\rightarrow 0$, $%
a\rightarrow 1$.

\subsection{Weak group entropies}

\label{sec:25}

We wish to mention that the same group-theoretical approach can also be
adopted for defining a huge class of \textit{trace-form} generalized
entropies. Precisely, in \cite{PT2016AOP} it has been shown that by
requiring that the composability axiom be satisfied on the uniform
distribution only (weak composability), one is lead to the expression of the 
\textit{universal-group entropy}, related with the universal formal group of
algebraic topology. This generalized entropy includes very many of the
trace-form entropies defined in the literature in the last decades. It
represents a weak formulation of the notion of group entropy, which can
still be used for many information-theoretical purposes (see for instance
the recent works \cite{GPB2020PHA}, \cite{HPPP2022ENT}).

\subsection{Outlook of group entropies}

\label{sec:26} To conclude this introduction to the theory of group
entropies with some concrete applications, we mention two recent research
lines.

\subsubsection{Group entropies and information geometry}

\label{sec:261}

Information geometry \cite{AmaMe,AmaIG} provides a new methodology whose
main objective is to investigate the properties of a statistical model by
means of underlying geometrical structures. In particular, one defines a
Riemannian metric in a manifold of probability distributions, together with
dually coupled affine connections. This geometric approach is especially
relevant in statistical inference, quantum information theory, machine
learning, convex optimization, time series analysis, etc. Here we will focus
on the concept of \emph{divergence} (and its associated geometric
structure), representing a pseudo-distance over the probability manifold.

There is a direct application of group entropy in the context of information
geometry. Indeed, a widely generalized, relative-entropy version of group
entropies, the $(h,f)$-\textit{divergence}, was introduced in \cite%
{RRT2019PRA}, in the context of information theory. Given two probability
distributions $p=(p_{1},...,,p_{W})$ and $q=(q_{1},...,,q_{W})$ such that $%
p_{i},q_{i}>0$, $i=1,\ldots ,W$, we can immediately deduce the \textit{%
relative $Z$-class}: 
\begin{equation}
Z_{G,\alpha }(p||q):=\log _{G}\Bigg[\Bigg(\sum_{i=1}^{W}p_{i}^{\alpha
}q_{i}^{1-\alpha }\Bigg)^{\frac{1}{\alpha -1}}\Bigg]\ ,  \label{eq:relGE}
\end{equation}%
$\alpha \neq 1$, which is nothing but a special version of the $(h,f)$%
-divergence, for $h(x)=G(\ln x^{\gamma })$, $f(x)=x^{\alpha }$, $\gamma =%
\frac{1}{\alpha -1}$. Clearly, alternative classes are easily obtainable
from the $(h,f)$-relative entropies of \cite{RRT2019PRA} by means of
different choices of $h$ and $f$. More general classes of group entropies
have been defined in \cite{RRT2019PRA}, \cite{PT2020CHA} and \cite{TJ2020SR}.

\subsubsection{Group entropies and negativity}

\label{sec:262}

The notion of logarithmic negativity introduced by Vidal and Werner in \cite%
{VW2002PRA} can also be generalized by means of a mathematical formalism
based on formal group theory. The main result is that there exists a
\textquotedblleft tower\textquotedblright\ of new, parametric information
measures, each of them reducing to the logarithmic negativity in a suitable
regime \cite{CMT2021QIP}.

\subsubsection{Multivariate family}

\label{sec:263}

Finally, we mention that a generalization of the class of $Z$-entropies has
been proposed in \cite{PT2020CHA}, where the notion of \textit{multivariate
group entropies} has been defined and several examples discussed.

\section{Permutation complexity}

\label{sec:3}

Henceforth we focus on random or deterministic, real-valued discrete time
processes that, in general, we simply call \textquotedblleft
processes\textquotedblright\ and denote by $\mathbf{X}=(X_{t})_{t\geq 0}$.
By a deterministic process we mean here that any realization $(x_{t})_{t\geq
0}$ of $\mathbf{X}$ is the orbit of a one dimensional dynamical system $(I,%
\mathcal{B},\mu ,f)$, where $I$ (the \textit{state space}) is a closed
interval of $\mathbb{R}$, $\mathcal{B}$ is the Borel $\sigma $-algebra of $I$%
, $\mu $ is a \textit{probability measure} over the measurable space $(I,%
\mathcal{B})$ (i.e., $\mu (I)=1$) and, for the time being, $f:I\rightarrow I$
is any $\mu $-invariant map (i.e., $\mu (f^{-1}(B))=\mu (B)$ for all $B\in 
\mathcal{B}$); alternatively, we say that $\mu $ is $f$-invariant. In the
deterministic case we will sometimes write $\mathbf{X}=(f^{t})_{t\geq 0}$,
where $f^{0}(x_{0})=x_{0}$ and $f^{t}(x_{0})=f(f^{t-1}(x_{0}))$, so that $%
(x_{t})_{t\geq 0}=(f^{t}(x_{0}))_{t\geq 0}$.

\subsection{Ordinal patterns}

\label{sec:31}

Let $x_{t}\in \mathbb{R}$ be an outcome of the process$\ \mathbf{X}$ at time 
$t$, i.e., an outcome of the univariate random variable $X_{t}$ or, in the
deterministic case, the point $f^{t}(x_{0})$ in the orbit of the initial
condition $x_{0}\in I$. More generally, given the integer $L\geq 2$, let the
string (window, block,...) $x_{t}^{L}:=x_{t},x_{t+1},\ldots ,x_{t+L-1}$ be
an outcome of the \textquotedblleft finite process\textquotedblright\ $%
X_{t}^{L}:=X_{t},X_{t+1},...,X_{t+L-1}$ or the orbit segment $%
(f^{n}(x_{0}))_{t\leq n\leq t+L-1}$ if $\mathbf{X}=(f^{t})_{t\geq 0}$. We
say that $x_{t}^{L}$ defines the \textit{ordinal pattern of length }$L$ (or 
\textit{ordinal }$L$\textit{-pattern} for short) $\mathbf{r}=(\rho _{0},\rho
_{1},\ldots ,\rho _{L-1})=\mathbf{r}(x_{t}^{L})$, where $\{\rho _{0},\rho
_{1},\ldots ,\rho _{L-1}\}\in \{0,1,\ldots ,L-1\}$, if 
\begin{equation}
x_{t+\rho _{0}}<x_{t+\rho _{1}}<\ldots <x_{t+\rho _{L-1}}  \label{ord patt}
\end{equation}%
(other rules can also be found in the literature). Alternatively, $x_{t}^{L}$
is said to be of \textit{type} $\mathbf{r}=\mathbf{r}(x_{t}^{L})$. In case
of two or more ties, one can adopt some convention, e.g., the earlier entry
is smaller. For example, if $L=4$ and 
\begin{equation}
x_{t}=0.3,\;x_{t+1}=-0.5,\;x_{t+2}=1.2,\;x_{t+3}=0.7,  \label{Ex ord patt}
\end{equation}%
then $\mathbf{r}(x_{t}^{4})=(1,0,3,2)$.

Ordinal $L$-patterns can be identified with permutations of $\{0,1,\ldots
,L-1\}$, i.e., with elements of the symmetric group of degree $L$, $\mathcal{%
S}_{L}$, whose cardinality, $\left\vert \mathcal{S}_{L}\right\vert $, is $L!$%
. In time series analysis, the symbolic representation of a time series $%
(x_{t})_{t\geq 0}$ by means of the sequence $(\mathbf{r}_{t})_{t\geq 0}$,
where $\mathbf{r}_{t}$ is the type of the (sliding) window $x_{t}^{L}$, is
called an \textit{ordinal representation}. This kind of symbolic
representations have proved to be useful in time series analysis \cite%
{Zanin2012,Amigo2013,Amigo2015B}, especially because the blocks of variables 
$X_{t}^{L}$ can detect dependencies of range $L$ among the variables $%
X_{t+\tau }$, $0\leq \tau \leq L-1$ \cite{Bandt2019,Weiss2022}. Ordinal
representations of time series define the corresponding finite-state
processes taking values on $\mathcal{S}_{L}$. In the deterministic case, two
conjugate dynamical systems $(I_{1},\mathcal{B}_{1},\mu _{1},f_{1})$ and $%
(I_{2},\mathcal{B}_{2},\mu _{2},f_{2})$ such that the conjugacy $\varphi
:I_{1}\rightarrow I_{2}$ preserves order, will output the same symbolic
orbit $(\mathbf{r}_{t})_{t\geq 0}$ for corresponding orbits $(x_{t})_{t\geq
0}$ and $(\varphi (x_{t}))_{t\geq 0}$. Ordinal representations generalize to
algebraic representations, where the algebraic structure of the alphabet
(e.g., $\mathcal{S}_{L}$ endowed with the composition of permutations as a
product) can be exploited; see \cite{Amigo2012,Haruna2019} for some
references.

Deterministic processes have some peculiarities that deserve a separate
discussion. If $\mathbf{X}=(f^{t})_{t\geq 0}$, then Equation (\ref{ord patt}%
) becomes%
\begin{equation}
f^{t+\rho _{0}}(x_{0})<f^{t+\rho _{1}}(x_{0})<\ldots <f^{t+\rho
_{L-1}}(x_{0}).  \label{ord patt f}
\end{equation}%
Correspondingly, the sets%
\begin{equation}
P_{\mathbf{r}}=\{x\in I:\text{ }(x,\,f(x),...,\,f^{L-1}(x))\text{ is of type 
}\mathbf{r}\in \mathcal{S}_{L}\}  \label{P_r}
\end{equation}%
define a finite partition of the state space $I$, 
\begin{equation}
\mathcal{P}_{L}=\{P_{\mathbf{r}}\neq \emptyset :\mathbf{r}\in \mathcal{S}%
_{L}\},  \label{Partition PL}
\end{equation}%
called the \textit{ordinal partition of order} $L$.

\begin{figure*}[tbp]
\begin{center}
\includegraphics[scale=0.43]{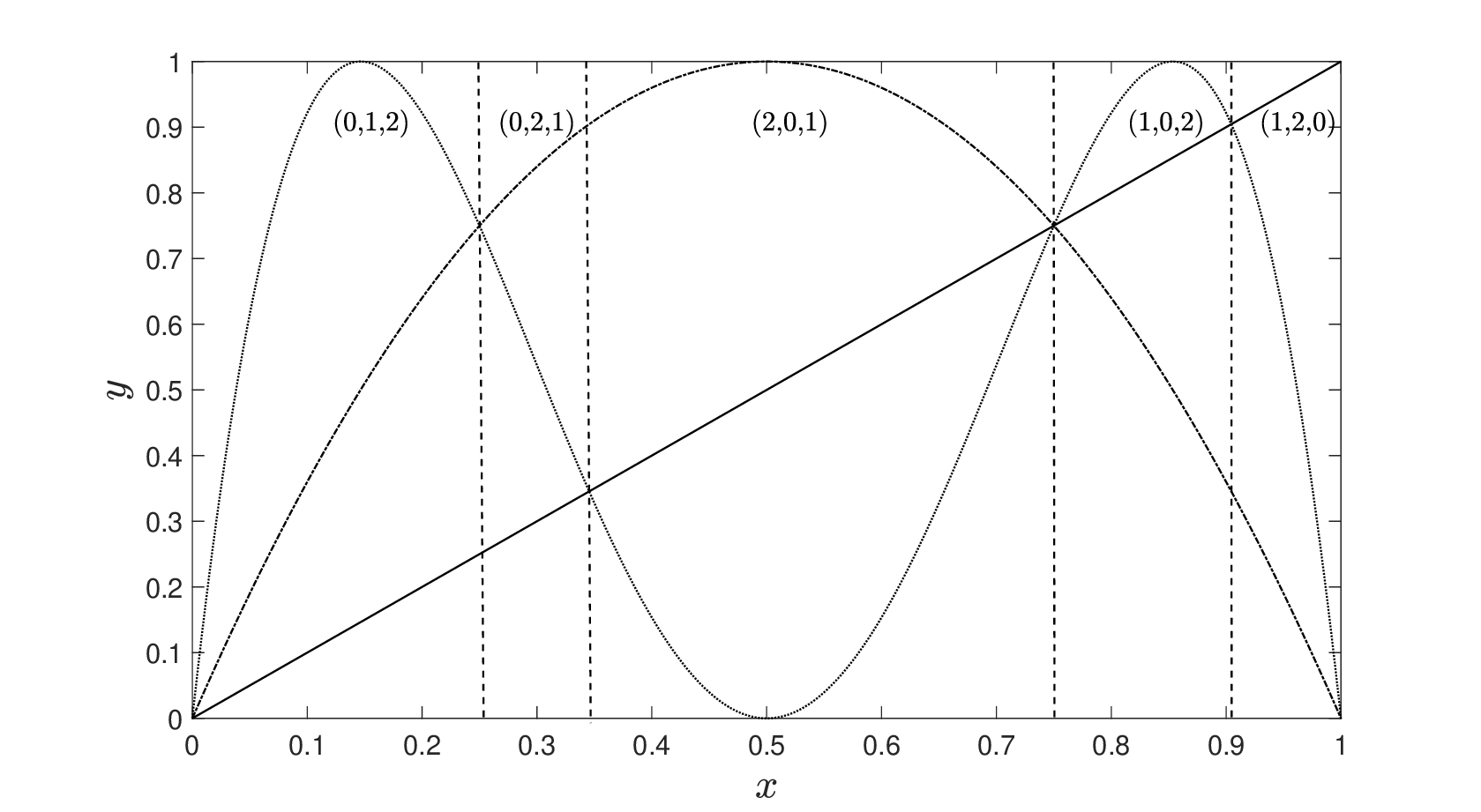}
\end{center}
\caption{Ordinal partition $\mathcal{P}_{3}$ of $[0,1]$ generated by the
ordinal 3-patterns of the full range logistic map $f(x)=4x(1-x)$. The solid
line (bisector) corresponds to $y=f^{0}(x)=x$, the dashed line to $y=f(x)$
and the dotted line to $y=f^{2}(x)$. With the rule \textquotedblleft earlier
is smaller\textquotedblright\ for the intersections of those lines, we find: 
$P_{(0,1,2)}=\left[ 0,\frac{1}{4}\right] \bigcup \left\{ \frac{3}{4}\right\} 
$, $P_{(0,2,1)}=\left( \frac{1}{4},\frac{5-\protect\sqrt{5}}{8}\right] $, $%
P_{(2,0,1)}=\left( \frac{5-\protect\sqrt{5}}{8},\frac{3}{4}\right) $, $%
P_{(1,0,2)}=\left( \frac{3}{4},\frac{5+\protect\sqrt{5}}{8}\right] $, $%
P_{(1,2,0)}=\left( \frac{5+\protect\sqrt{5}}{8},1\right] $ and $%
P_{(2,1,0)}=\emptyset .$}
\label{fig:Figura1}
\end{figure*}

For moderate $L$'s, the sets $P_{\mathbf{r}}$ can be found graphically \cite%
{A2010BOOK}. For example, Figure~\ref{fig:Figura1} shows the partition $%
\mathcal{P}_{3}$ for the full range logistic map 
\begin{equation}
f(x)=4x(1-x),\;0\leq x\leq 1.  \label{logistic map}
\end{equation}
Notice that no point has type $(2,1,0)$, i.e., there is no $x\in \lbrack
0,1] $ such that $f^{2}(x)<f(x)<x$ and, hence, $P_{(2,1,0)}=\emptyset $.

Coarse-graining of the state space of a measure-preserving dynamical system
by a finite partition induces in a standard way a stationary, finite-state
random process, called the \textit{symbolic dynamics} with respect to the
partition. We briefly remind its construction in the case of the ordinal
partition $\mathcal{P}_{L}$ of $I$. Define the measurable maps $X_{t}^{%
\mathcal{P}_{L}}:I\rightarrow \mathcal{S}_{L}$ by%
\begin{equation}
X_{t}^{\mathcal{P}_{L}}(x)=\mathbf{r}\;\;\Leftrightarrow \;\;f^{t}(x)\in P_{%
\mathbf{r}}\text{.}  \label{symb dynamics}
\end{equation}%
Then it is straighforward to show that $\mathbf{X}^{\mathcal{P}_{L}}=(X_{t}^{%
\mathcal{P}_{L}})_{t\geq 0}$ is a stationary random process \cite{A2010BOOK}%
. In particular, the probability $p(\mathbf{r})$ that $X_{t}^{\mathcal{P}%
_{L}}=\mathbf{r}\in \mathcal{S}_{L}$, i.e., the probability that $%
X_{t}^{L}=x_{t}^{L}$ is of type $\mathbf{r}\in \mathcal{S}_{L}$ is given by 
\begin{equation}
p(\mathbf{r})=\mu (P_{\mathbf{r}}).  \label{p(r)}
\end{equation}%
Note that the ordinal representation $(\mathbf{r}_{t})_{t\geq 0}$ of the
orbit $(f^{t}(x_{0}))_{t\geq 0}=(x_{t})_{t\geq 0}$ is nothing other than the
trajectory $\mathbf{X}^{\mathcal{P}_{L}}(x_{0})$, i.e., the realization of a
stationary, finite-state random process. In fact, while $x_{t}=x$ implies $%
x_{t+1}=f(x)$ (so that knowing the initial condition $x_{0}$ is equivalent
to knowing the complete deterministic orbit $(f^{t}(x_{0})_{t\geq 0})$), $%
\mathbf{r}_{t}=\mathbf{r}\in \mathcal{S}_{L}$ implies $\mathbf{r}_{t+1}=%
\mathbf{r}^{\prime }\in \mathcal{S}_{L}$ only with probability%
\begin{equation}
p(X_{t+1}^{\mathcal{P}_{L}}=\mathbf{r}^{\prime }\left\vert X_{t}^{\mathcal{P}%
_{L}}=\mathbf{r}\right) =\frac{\mu (P_{\mathbf{r}}\cap f^{-1}P_{\mathbf{r}%
^{\prime }})}{\mu (P_{\mathbf{r}})}.  \label{cond prob}
\end{equation}

\begin{example}
As way of illustration, consider again the full range logistic map (\ref%
{logistic map}) and the ordinal partition\ $\mathcal{P}_{6}$ of $I=[0,1]$;
see the caption of Figure \ref{fig:Figura1} for the intervals $P_{\mathbf{r}%
}\in \mathcal{P}_{6}$. Using the natural invariant measure \cite{Eckmann1985}
\begin{equation}
\mu (P_{\mathbf{r}})=\int_{P_{\mathbf{r}}}\frac{dx}{\pi \sqrt{x(1-x)}},
\label{nat measure}
\end{equation}%
we obtain%
\begin{equation*}
\begin{array}{l}
\mu (P_{(0,1,2)})=\mu ([0,\frac{1}{4}])=\frac{10}{30} \\ 
\mu (P_{(1,0,2)}\cap f^{-1}P_{(0,1,2)})=\mu ([0,\frac{1}{4}(2-\sqrt{3})])=%
\text{ }\frac{5}{30} \\ 
\mu (P_{(1,0,2)}\cap f^{-1}P_{(0,2,1)})=\mu ([\frac{1}{4}(2-\sqrt{3}),\frac{1%
}{8}(3-\sqrt{5})])=\frac{1}{30} \\ 
\mu (P_{(1,0,2)}\cap f^{-1}P_{(2,0,1)})=\mu ([\frac{1}{8}(3-\sqrt{5}),\frac{1%
}{4}])=\frac{4}{30}%
\end{array}%
\end{equation*}%
and $P_{(1,0,2)}\cap f^{-1}P_{\mathbf{r}^{\prime }}=\emptyset $ for $\mathbf{%
r}^{\prime }=(1,0,2),(1,2,0),(2,1,0)$. Apply now Equation (\ref{cond prob})
to calculate the conditional probabilities $p(\mathbf{r}^{\prime }\left\vert
(0,1,2)\right) :=p(X_{t+1}^{\mathcal{P}_{3}}=\mathbf{r}^{\prime }|X_{t}^{%
\mathcal{P}_{3}}=(0,1,2))$, with the following results: 
\begin{eqnarray*}
p((0,1,2)\left\vert (0,1,2)\right) &=&0.5,\;p((0,2,1)\left\vert
(0,1,2)\right) =0.1, \\
p((2,0,1)\left\vert (0,1,2)\right) &=&0.4
\end{eqnarray*}%
and $p(\mathbf{r}^{\prime }\left\vert (0,1,2)\right) =0$ otherwise.
Therefore, if $(\mathbf{r}_{t})_{t\geq 0}$ is the ordinal representation of
an orbit of the full range logistic map, then the pattern $(0,1,2)$ is
followed by $(0,1,2)$ in $50\%$ of the cases, by $(0,2,1)$ in $10\%$ of the
cases and $(2,0,1)$ in $40\%$ of the cases. Other \textquotedblleft
transition probabilities\textquotedblright\ $p(\mathbf{r}^{\prime
}\left\vert \mathbf{r}\right) $ are calculated in a similar way.
\end{example}

Since we wish to deal with deterministic and random processes on an equal
footing and, moreover, stationarity is necessary also for random processes
regarding some theoretical and practical properties below, we assume the
following weaker form of stationarity henceforth.

\medskip

\textbf{Stationarity Condition} \cite{Bandt2002}. \textit{For }$k\leq L-1,$%
\textit{\ the probability for }$x_{t}<x_{t+k}$\textit{\ should not depend on 
}$t$\textit{.}

\medskip

Sometimes we call processes that fulfill the Stationary Condition \textit{%
weakly stationary}. They include not only the deterministic and stationary
random processes but also non-stationary random processes with stationary
increments such as the fractional Brownian motion \cite{Mandelbrot1968} and
its increments, that is, the fractional Gaussian noise. We will use these
random processes, which have long range dependencies, in some numerical
examples.

\subsection{Asymptotic growth of allowed patterns}

\label{sec:32}

An ordinal pattern $\mathbf{r}\in \mathcal{S}_{L}$ is \textit{allowed} for
the finite process $X_{t}^{L}$ if the probability that a string $x_{t}^{L}$
of type $\mathbf{r}$ is output by $X_{t}^{L}$ is positive. Otherwise, the
ordinal $L$-pattern $\mathbf{r}$ is \textit{forbidden} for $X_{t}^{L}$.
Likewise, an ordinal pattern $\mathbf{r}\in \mathcal{S}_{L}$ is \textit{%
allowed} for the process $\mathbf{X}$ if the probability that a string $%
x_{t}^{L}$ of type $\mathbf{r}$ is output by $\mathbf{X}$ is positive, that
is, there is a $t\geq 0$ such that $x_{t}^{L}$ is allowed for $X_{t}^{L}$.
Otherwise $\mathbf{r}$ is \textit{forbidden} for $\mathbf{X}$. If $\mathbf{X}
$ is weakly stationary (as we assume unless otherwise stated), then $\mathbf{%
r}\in \mathcal{S}_{L}$ is allowed for $\mathbf{X}$ if and only if it is
allowed for any given $X_{t}^{L}$, since then the probabilities of the $L$%
-patterns do not depend on $t$.

Figure~\ref{fig:Figura1} shows that the ordinal $3$-pattern $\mathbf{r}%
=(2,1,0)$ is forbidden for the logistic map $f(x)=4x(1-x)$ or, equivalently, 
$P_{(2,1,0)}=\emptyset $; all other $3$-patterns, are allowed. More
generally, the number of allowed $L$-patterns for a deterministic process is
simply $\left\vert \mathcal{P}_{L}\right\vert $, the cardinality of ordinal
partition of order $L$ of the state space, see (\ref{Partition PL}). So,
according to Figure~\ref{fig:Figura1}, $\left\vert \mathcal{P}%
_{3}\right\vert =5$ is the number of allowed $3$-patterns for the full range
logistic map.

\begin{remark}
A word of caution for the practitioners is in order at this point:
\textquotedblleft missing\textquotedblright\ patterns in numerical
simulations are not necessarily forbidden. Indeed, allowed patterns are
\textquotedblleft visible\textquotedblright\ in numerical computations only
if they belong to stable (sufficiently long) orbits. For example, in the
well-known bifurcation diagram of the logistic family of maps $%
f_{a}(x)=ax(1-x)$, $0\leq x\leq 1$, only a period-$3$ orbit appears in the
parametric window $3.83\lesssim a\lesssim 3.84$, in accordance with Singer's
Theorem on the stability of orbits of quadratic maps \cite{Singer1978}.
Hence, one might erroneously conclude that $f_{a}(x)$ has only three allowed 
$L$-patterns for every $L\geq 3$. However, according to Sarkovskii's Theorem 
\cite{Sarko1964}, for each $a$ in the period-$3$ window, there are orbits of
any possible period (\textquotedblleft period $3$ implies
chaos\textquotedblright\ \cite{LiYorke1975}). In fact, we will see shortly
that the number of allowed $L$-patterns grows exponentially with $L$ for
most interval maps of practical interest.
\end{remark}

Let $\mathcal{A}_{L}(\mathbf{X})$ denote the number of allowed $L$-patterns
for a process $\mathbf{X}$. Due to the Stationarity Condition, 
\begin{eqnarray}
\mathcal{A}_{L}(\mathbf{X}) &=&\left\vert \{\text{allowed }L\text{-patterns
for }\mathbf{X}\}\right\vert  \label{A_L(X)} \\
&=&\left\vert \{\text{allowed patterns for }X_{t}^{L}\}\right\vert  \notag
\end{eqnarray}%
for all $t$. Note that $\mathcal{A}_{L}(\mathbf{X})$ is a non-decreasing
function of $L$.

Let $I$ be a closed interval of $\mathbb{R}$. A map $f:I\rightarrow I$ is
called \textit{piecewise monotone} if there is a finite partition of $I$
into intervals such that $f$ is continuous and strictly monotone on each
(open, closed, closed from one side or degenerate) subinterval of the
partition. Most one-dimensional maps found in practice are piecewise
monotone, so this condition does not imply any strong restriction for
practical purposes. According to a theorem due to Bandt, Keller and Pompe 
\cite{Bandt2002B}, 
\begin{eqnarray}
&&\ln \mathcal{A}_{L}(\mathbf{X})  \label{allowed pat f} \\
&=&\ln \left\vert \{\text{allowed }L\text{-patterns for deterministic }%
\mathbf{X}\}\right\vert \sim h_{0}(f)L,  \notag
\end{eqnarray}%
where $f$ is the (piecewise monotone) map generating the outputs of $\mathbf{%
X}$, $h_{0}(f)$ is the topological entropy of $f$ \cite{Walters2000} and the
symbol $\sim $ stands for \textquotedblleft asymptotically when $%
L\rightarrow \infty $\textquotedblright\ (i.e., $\lim_{L\rightarrow \infty
}\ln \mathcal{A}_{L}(\mathbf{X})/(h_{0}(f)L)=1$). Therefore, the number of
allowed $L$-patterns for a piecewise monotone map $f$ with positive
topological entropy grows exponentially with $L$. To be more precise,
according to the proof of Proposition 1(b) in \cite{Bandt2002B}, $\mathcal{A}%
_{L}(\mathbf{X})\sim \exp [h_{0}(f)L+\ln L+\mathrm{const}]$. For this
reason, by a \textit{deterministic process} we mean here and hereafter the
dynamics generated by a chaotic multi-humped interval map, so that Equation (%
\ref{allowed pat f}) holds with $h_{0}(f)>0$. Such is the case of the
logistic family, $f_{a}(x)=ax(1-x)$ for $a>3.5699456...$ (the Feigenbaum
point) and the usual \textquotedblleft chaotic\textquotedblright\ maps.

Since, on the other hand, the number of possible $L$-patterns is $L!$ and 
\begin{equation}
\ln L!\sim L\ln L  \label{Stirling}
\end{equation}%
by Stirling's formula $\ln L!\simeq L(\ln L-1)+\tfrac{1}{2}\ln (2\pi L)$, we
conclude from Equation (\ref{allowed pat f}) that deterministic processes
necessarily have forbidden $L$-patterns for $L$ large enough and, in fact,
the number of forbidden $L$-patterns of deterministic processes grows
super-exponentially with $L$. For the structure and growth with $L$ of
forbidden patterns in deterministic processes, as well as their application
to the discrimination of deterministic signals from random signals, the
interested reader is referred to \cite{A2010BOOK} and references therein.
Forbidden patterns there exist in higher dimensional dynamics too, where
they can be defined in different ways \cite{Amigo2013B}. Therefore, we
expect forbidden patterns in the projections of higher dimensional orbits,
as shown numerically in \cite{Amigo2008} using lexicographical order.

At the other extreme are random processes without forbidden patterns, that
is, processes for which all ordinal patterns of any length are allowed and,
hence, their growth is factorial: $\mathcal{A}_{L}(\mathbf{X})=\left\vert 
\mathcal{S}_{L}\right\vert =L!$. A trivial example of a random process
without forbidden patterns is white noise.

Also \textit{noisy} deterministic time series may not have forbidden
patterns (for sufficiently long series). Indeed, when the dynamics takes
place on a nontrivial attractor so that the orbits are dense, then the
observational (white) noise will \textquotedblleft
destroy\textquotedblright\ all forbidden patterns in the long run, no matter
how small the noise amplitude. For this reason, we sometimes call noisy
deterministic processes and weakly stationary random processes without
forbidden patterns just \textit{forbidden-pattern-free }(FPF) \textit{%
processes} or signals. In this case, 
\begin{eqnarray}
\ln \mathcal{A}_{L}(\mathbf{X}) &=&\ln \left\vert \{\text{allowed }L\text{%
-patterns for FPF }\mathbf{X}\}\right\vert  \label{allowed pat X} \\
&=&\ln L!\sim L\ln L  \notag
\end{eqnarray}%
unlike Equation (\ref{allowed pat f}), where we used the asymptotic
equivalence (\ref{Stirling}). Since the real world is allegedly nonlinear
and noisy, one expects a factorial growth \ of allowed patterns in empirical
observations.

\begin{figure*}[tbp]
\begin{center}
\includegraphics[scale=0.43]{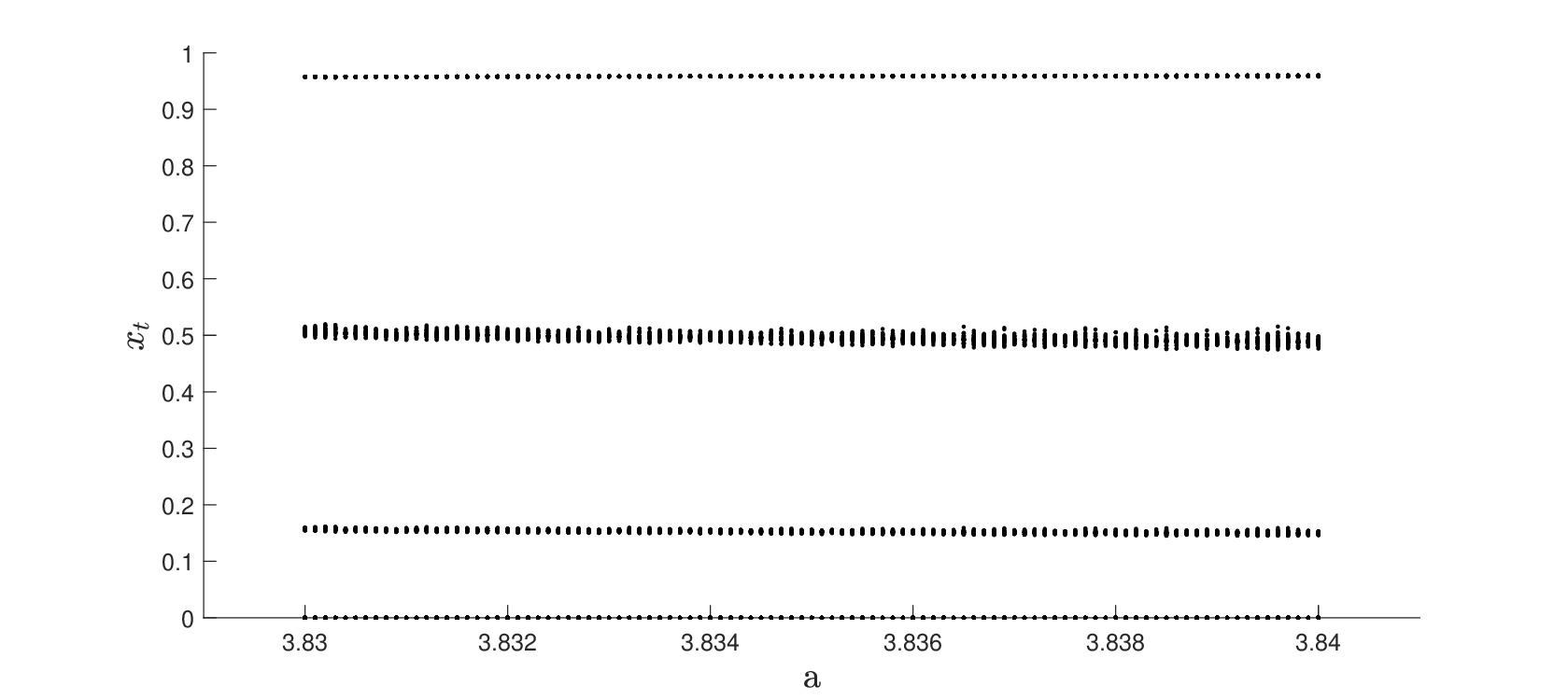}
\end{center}
\caption{ Example of a random process with forbidden patterns: the logistic
parabola $f_{a}(x)=ax(1-x)$ with $3.83\leq a\leq 3.84$ (period-3 window) and
additive white noise of amplitude $0.002$, see Equation (\protect\ref%
{dynamical noise}).}
\label{fig:Figura2}
\end{figure*}

To complete the picture, let us point out that random processes can have
forbidden patterns too. Think of a periodic signal with observational noise
such that the noise bands do not overlap. The same can happen with dynamical
noise if the orbit is stable and the noise amplitude is sufficiently small.
As way of illustration, Figure~\ref{fig:Figura2} shows the orbits of the
logistic map with additive noise, 
\begin{equation}
x_{t+1}=ax_{t}(1-x_{t})+\varepsilon _{t}\,,  \label{dynamical noise}
\end{equation}%
where $3.83\leq a\leq 3.84$ (period-3 window) and the random variables $%
\varepsilon _{t}$ are independent and uniformly distributed in $%
[-0.001,0.001]$; the only allowed 3-patterns of this noisy map are $(0,1,2)$
and its cyclic permutations $(1,2,0)$ and $(2,0,1)$. Toy models of random
processes with subfactorial growths, specifically $\ln \mathcal{A}_{L}(%
\mathbf{X})$ $\sim cL\ln L$ with $0<c<1$, were constructed in \cite%
{ADT2022CNS}.

\subsection{Permutation complexity functions and classes}

\label{sec:33}

In Section \ref{sec:24} we discussed an approach to measure the complexity
of statistical systems in the form of extensive group entropies, based on
universality classes (sub-exponential, exponential and super-exponential).
To follow this approach, we capitalize on $\mathcal{A}_{L}(\mathbf{X})$,
which is here the counterpart of the function $\mathcal{W}(N)$ there, $L$
playing here the role of $N$ (the \textquotedblleft
extensive\textquotedblright\ parameter).

\begin{definition}
\label{DefClass}Let $g(t)$ be a positive, invertible and sufficiently
regular function of the real variable $t\geq 0$. A process $\mathbf{X}$ is
said to belong to the permutation complexity (PC) class $g$ if%
\begin{equation}
\ln \mathcal{A}_{L}(\mathbf{X})\sim g(L)  \label{pcf}
\end{equation}%
as $L\rightarrow \infty $.
\end{definition}

Correspondingly, the function $g(t)$ will be called the \textit{permutation
complexity} (PC)\textit{\ function} of the process $\mathbf{X}$.

\begin{remark}
\label{Remark PCF}Two important observations on the PC function of a process:

\begin{enumerate}
\item Regarding regularity, we will assume henceforth that $g(t)$ is
bicontinuous, i.e., both $g(t)$ and its inverse $g^{-1}(s)$ are continuous.
The bicontinuity and invertibility of $g(t)$ imply that $g(t)$ and, hence, $%
g^{-1}(t)$ are strictly monotonic \cite{Apostol1974}, in fact, strictly
increasing in our case. The differentiability of $g$ will be considered in
Remark \ref{Differentiability_g}.

\item Regarding uniqueness, the complexity class $g$ depends only on the
asymptotic behavior of $g(t)$; any other function $\tilde{g}(t)\sim g(t)$
(i.e., $\tilde{g}(t)=g(t)+o(g(t))$) will work out as well. Put in other
terms, PC classes are defined up to asymptotic equivalence.
\end{enumerate}
\end{remark}

We have already met two distinct PC functions: for deterministic processes, $%
\ln \mathcal{A}_{L}(\mathbf{X})$ is linear in $L$ (see Equation (\ref%
{allowed pat f})), while for FPF processes, $\ln \mathcal{A}_{L}(\mathbf{X})$
is super-linear in $L$ (see Equation (\ref{allowed pat X})). This prompts
the following classification of processes into PC classes:

\begin{description}
\item[(C1)] \textit{Exponential class}: $\ln \mathcal{A}_{L}(\mathbf{X})\sim
cL$ ($c>0$), i.e.,%
\begin{equation}
g(t)=ct=:g_{\text{exp}}(t).  \label{g_exp}
\end{equation}

\item[(C2)] \textit{Factorial class}: $\ln \mathcal{A}_{L}(\mathbf{X})\sim
L\ln L$, i.e., 
\begin{equation}
g(t)=t\ln t=:g_{\text{fac}}(t).  \label{g_fact}
\end{equation}

\item[(C3)] \textit{Sub-factorial (or super-exponential) class}: $\ln 
\mathcal{A}_{L}(\mathbf{X})\sim g_{\text{sub}}(t)$, where (i) $g_{\text{exp}%
}(t)=o(g_{\text{sub}}(t))$ and $g_{\text{sub}}(t)=o(g_{\text{fac}}(t))$ or,
else, (ii)%
\begin{equation}
g_{\text{sub}}(t)=ct\ln t\;\text{\ with\ \ }0<c<1.  \label{gamma_sub}
\end{equation}
\end{description}

\begin{remark}

\begin{enumerate}
\item Any function of the form $g(t)=ct$ is the PC function of a
deterministic process generated by a piecewise linear map of the interval $%
[0,1]$ with constant slopes $\pm e^{c}$.

\item Examples of functions $g_{\text{sub}}$ satisfying condition C3(i) for
the subfactorial class are 
\begin{equation}
g_{\text{\emph{sub}}}(t)=t\ln ^{(n)}t\;\;(n\geq 2),  \label{gamma_sub2}
\end{equation}%
where $\ln ^{(n)}t$ denotes the composition of the logarithmic function $n$
times.
\end{enumerate}
\end{remark}

\bigskip

In sum, there are two main PC classes: the exponential class, populated by
deterministic (noiseless) processes, and the factorial class, populated by
the \textquotedblleft usual\textquotedblright\ random processes, including
noisy deterministic processes. Such processes can be part of theoretical
models (where $\mathcal{A}_{L}(\mathbf{X})$ and, hence, $g(t)$ can be
derived from the model) or the result of practical applications such as
numerical simulations (exponential or factorial class), analysis of denoised
nonlinear signals (exponential class) and analysis of real world time series
(factorial class). In contrast, sub-factorial processes are hard to find;
toy models with subfactorial PC functions of the form (\ref{gamma_sub}) were
presented in \cite{ADT2022CNS}. Finally, one dimensional dynamical systems
with zero topological entropy (think of a periodic dynamic) would belong to
a sub-exponential class, while there is no super-factorial PC classes with
the convention for ordinal patterns that we use.

\begin{figure*}[tbp]
\begin{center}
\includegraphics[scale=0.43]{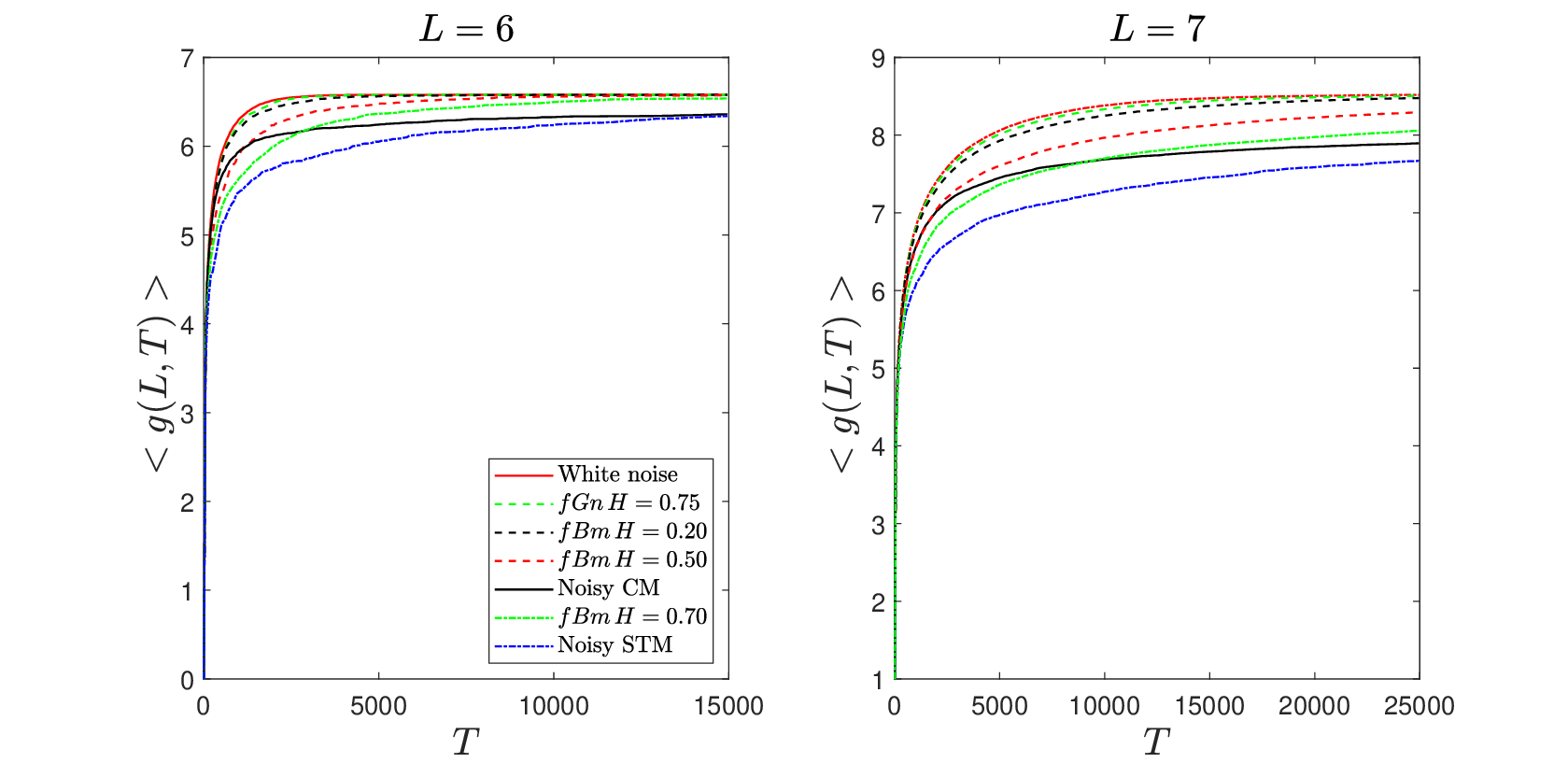}
\end{center}
\caption{Averages of the finite PC function $g(L,T)$ over 10 realizations of
the random processes listed in the inset are plotted vs $T$ (the time series
length) for $L=6$ and $6\leq T\leq 15,000$ (left panel) and $L=7$ and $7\leq
T\leq 25,000$ (right panel). See the text for details.}
\label{fig:Figura3}
\end{figure*}

Figure~\ref{fig:Figura3} shows the \textit{finite PC function} $g(L,T)$,
i.e., the logarithm of the number of allowed $L$-patterns in time series of
length $T$ with (left panel) $L=6$, and $6\leq T\leq 15,000$, as well as
(right panel) $L=7$, and $7\leq T\leq 25,000$, averaged over 10 realizations
for the following FPF processes: (\textbf{P1}) \textit{White noise} (WN) in
the form of an independent and uniformly distributed process on $[0,1]$; (%
\textbf{P2}) \textit{Fractional Gaussian noise} (fGn) with Hurst exponent $%
H=0.75$ \cite{Mandelbrot1968}; (\textbf{P3}) \textit{Fractional Brownian
motion} (fBm) with $H=0.2$ (anti-persistent process), $H=0.5$ (classical
Brownian motion) and $H=0.70$ (persistent process) \cite{Mandelbrot1968}; (%
\textbf{P4}) \textit{Cubic map} \textit{with observational white noise} of
amplitude $0.15$ (Noisy CM), i.e., $x_{t}=y_{t}+z_{t}$, where $%
y_{t}=3y_{t-1}(1-y_{t-1}^{2})$ and $(z_{t})_{t\geq 0}$ is WN with $%
\left\vert z_{t}\right\vert \leq 0.075$; (\textbf{P5}) \textit{Skew tent map}
\textit{with observational white noise} of amplitude $0.20$ (Noisy STM),
i.e., $x_{t}=y_{t}+z_{t}$, where $y_{t}=y_{t-1}/0.25$ if $0\leq y_{t-1}\leq
0.25$ and $y_{t}=(1-y_{t-1})/0.75$ if $0.25\leq y_{t-1}\leq 1$, and $%
(z_{t})_{t\geq 0}$ is WN with $\left\vert z_{t}\right\vert \leq 0.10$. For
calculation purposes, the maximal length of the time series was set at $%
T_{\max }=50,000$ ($\gg 7!=5,040$), but the computational loop is actually
exited as soon as the probability distribution of the $L$-patterns
stabilizes. Although 
\begin{equation*}
\lim_{T\rightarrow \infty }g(L,T)=\ln L!=\left\{ 
\begin{array}{cc}
6.\,\allowbreak 579\,3 & \text{for }L=6 \\ 
8.\,\allowbreak 525\,2 & \text{for }L=7%
\end{array}%
\right.
\end{equation*}%
for all processes (P1)-(P5) (see Equation (\ref{allowed pat X})), the
convergence of the two noisy deterministic processes (P4) and (P5) is
remarkably slow due to the small amplitude of the WN contamination. The
different growth rates of $g(L,T)$ can be used to discriminate FPF processes 
\cite{ADT2022CNS}. Of course, WN is the process that materializes all $L!$
allowed $L$-patterns fastest.

\section{Group entropy meets permutation complexity}

\label{sec:4}

Let $p(\mathbf{r})$ be the probability that the type of $X_{t}^{L}=x_{t}^{L}$
is the ordinal pattern $\mathbf{r}\in \mathcal{S}_{L}$. Then, the
conventional (metric) permutation entropy of the finite process $X_{t}^{L}$, 
$H^{\ast }(X_{t}^{L})$, is defined as the Shannon entropy of the probability
distribution $\{p(\mathbf{r}):\mathbf{r}\in \mathcal{S}_{L}\}$, i.e., 
\begin{equation}
H^{\ast }(X_{t}^{L})=-\sum_{\mathbf{r}\in \mathcal{S}_{L}}p(\mathbf{r})\ln p(%
\mathbf{r}),  \label{h*_mu,L}
\end{equation}%
while the topological permutation entropy of the finite process $X_{t}^{L}$
is defined as 
\begin{equation}
\;H_{0}^{\ast }(X_{t}^{L})=\ln \mathcal{A}_{L}(\mathbf{X}),  \label{h*_0,L}
\end{equation}%
see Equation (\ref{A_L(X)}). From these definitions it follows that%
\begin{equation}
H^{\ast }(X_{t}^{L})\leq H_{0}^{\ast }(X_{t}^{L})\leq \ln L!,  \label{bounds}
\end{equation}%
where $H^{\ast }(X_{t}^{L})=H_{0}^{\ast }(X_{t}^{L})$ for uniform
probability distributions of the allowed $L$-patterns, and $H_{0}^{\ast
}(X_{t}^{L})=\ln L!$ if all $L$-patterns are allowed. The convenience of
denoting the topological version of permutation entropy by a subscript $0$
will become clear later.

Remarkably, if $\mathbf{X}$ is the deterministic process generated by a
(piecewise monotone) map $f$ with Kolmogorov-Sinai entropy $h(f)$ \cite%
{Walters2000}, then the metric permutation entropy (rate) of $\mathbf{X}$,%
\begin{equation}
h^{\ast }(\mathbf{X})=\underset{L\rightarrow \infty }{\lim \sup }\frac{1}{L}%
H^{\ast }(X_{t}^{L}),  \label{h*_mu}
\end{equation}%
converges to $h(f)$, the Kolmogorov-Sinai entropy of $f$: $h^{\ast }(\mathbf{%
X})=h(f)$ \cite{Bandt2002B}. However, $h^{\ast }(\mathbf{X})$ cannot be
extended from deterministic to random processes. Indeed, the topological
permutation entropy (rate) of a process $\mathbf{X}$, 
\begin{equation}
h_{0}^{\ast }(\mathbf{X})=\underset{L\rightarrow \infty }{\lim \sup }\frac{1%
}{L}H_{0}^{\ast }(X_{t}^{L})=\underset{L\rightarrow \infty }{\lim \sup }%
\frac{1}{L}\ln \mathcal{A}_{L}(\mathbf{X}),  \label{h*_0}
\end{equation}%
diverges as soon as $\mathcal{A}_{L}(\mathbf{X})$ grows super-exponentially
with $L$. In particular, for FPF processes 
\begin{equation}
h_{0}^{\ast }(\mathbf{X})=\underset{L\rightarrow \infty }{\lim \sup }\frac{1%
}{L}\ln \mathcal{A}_{L}(\mathbf{X})=\lim_{L\rightarrow \infty }\ln L=\infty
\label{h*_0(WN)}
\end{equation}%
by Equation (\ref{allowed pat X}). Therefore, we cannot expect the metric
permutation entropy of noisy processes to converge either. Note that the
entropy rates $h^{\ast }(\mathbf{X})$ and $h_{0}^{\ast }(\mathbf{X})$ do not
depend on the initial time $t$; initial times $t\gg 0$ are usual in time
series analysis to eliminate possible transients.

In this section we use the extensivity of the $Z$-entropy (Section \ref%
{sec:24}), this time applied to the permutation complexity classes (Section %
\ref{sec:33}) to extend conventional permutation entropy $h_{0}^{\ast }(%
\mathbf{X})$ from deterministic processes to random processes in such a way
that the \textquotedblleft topological permutation group
entropy\textquotedblright\ $z_{0}^{\ast }(\mathbf{X})$ converges for all
processes in the same complexity class as $\mathbf{X}$ and, moreover, $%
z_{0}^{\ast }(\mathbf{X})=h_{0}^{\ast }(\mathbf{X})$ in the exponential
class. Indeed, as we will see below, $z_{0}^{\ast }(\mathbf{X})<\infty $
amounts to the extensitivity of the $Z$-entropy.

\subsection{Permutation entropy of finite order}

\label{sec:41}

To adapt the concept of $Z$-entropy to the context of permutation
complexity, let us remind that $\mathcal{A}_{L}(\mathbf{X})$ plays here the
role of $\mathcal{W}(N)$ in Theorem \ref{theo1} and $\ln \mathcal{A}_{L}(%
\mathbf{X})=g(L)$ asymptotically. If we substitute $\mathcal{W}(t)=e^{g(t)}$
(i.e., $\mathcal{W}^{-1}(s)=g^{-1}(\ln s)$) in Equation (\ref{maineq}) and
dispense with the factor 
\begin{equation}
\lambda =\frac{1}{(\mathcal{W}^{-1})^{\prime }(1)}=\mathcal{W}^{\prime }(%
\mathcal{W}^{-1}(1))=g^{\prime }(g^{-1}(0))  \label{lambda}
\end{equation}%
(which would require $g$ to be differentiable), we obtain the following
definition.

\begin{definition}
Let $g(t)$ be the PC function of a process $\mathbf{X}$. The \emph{(metric)
permutation entropy of order} $L$ of $\mathbf{X}$ is defined as%
\begin{equation}
Z_{g,\alpha }^{\ast }(X_{t}^{L})\equiv Z_{g,\alpha }^{\ast
}(p)=g^{-1}(R_{\alpha }(p))-g^{-1}(0),  \label{Zg_entropy}
\end{equation}%
where $\alpha >0$, $p$ is the probability distribution of the ordinal $L$%
-patterns of $X_{t}^{L}=X_{t},X_{t+1},...,X_{t+L-1}$, and $R_{\alpha }(p)$
is R\'{e}nyi's entropy (\ref{Renyi ent}).
\end{definition}

Remember that, according to Remark \ref{Remark PCF}, $g^{-1}(s)$ is
continuous and strictly increasing. Note that if $R_{\alpha }(p)=0$, then $%
Z_{g,\alpha }^{\ast }(p)=0$ too. By its definition (and the increasing
monotonicity of $g^{-1}(s)$), $Z_{g,\alpha }^{\ast }(X_{t}^{L})$ inherits
some of the properties of $R_{\alpha }(p)$. For instance, $Z_{g,\alpha
}^{\ast }(X_{t}^{L})$ is monotone decreasing with respect to the parameter $%
\alpha $ \cite{Amigo2018},%
\begin{equation}
Z_{g,\alpha }^{\ast }(X_{t}^{L})\geq Z_{g,\beta }^{\ast }(X_{t}^{L})\;\;%
\text{for\ \ }\alpha <\beta  \label{hierarchy}
\end{equation}%
and each $L\geq 2$.

For the exponential class, we know from Section \ref{sec:24}, Equation (\ref%
{Renyi ent}), that $Z_{g,\alpha }^{\ast }(p)$ is the R\'{e}nyi entropy. The
next theorem gives $Z_{g,\alpha }^{\ast }(p)$ also for the factorial and
subfactorial classes. For the factorial and sub-factorial classes (the
latter with $g_{\text{\textrm{sub}}}(t)=ct\ln t$, $0<c<1$), the principal
branch of the Lambert function $\mathcal{L}(x)$ (Section \ref{sec:24},
Equation (\ref{JTTP})) is needed. Complexity functions of the form (\ref%
{gamma_sub2}) require a generalization of $\mathcal{L}(x)$ \cite{ADT2022CNS}.

\begin{theorem}
\label{ThmZ_class}Let $p$ be the probability distribution of the ordinal $L$%
-patterns of $X_{t}^{L}$. For the PC classes (C1)-(C3) of Section \ref%
{sec:33}, the following holds.

\begin{description}
\item[(a)] For $g_{\text{\emph{exp}}}(t)=ct$:%
\begin{equation}
Z_{g_{\text{\emph{exp}}},\alpha }^{\ast }(X_{t}^{L})=\frac{1}{c}R_{\alpha
}(p)=:Z_{\text{\emph{exp}},\alpha }^{\ast }(X_{t}^{L}).  \label{Z_exp}
\end{equation}

\item[(b)] For $g_{\text{\emph{fac}}}(t)=t\ln t$:%
\begin{equation}
Z_{g_{\text{\emph{fac}}},\alpha }^{\ast }(X_{t}^{L})=e^{\mathcal{L}%
[R_{\alpha }(p)]}-1=:Z_{\text{\emph{fac}},\alpha }^{\ast }(X_{t}^{L}).
\label{Z_fac}
\end{equation}

\item[(c)] For $g_{\text{\emph{sub}}}(t)=ct\ln t$ $(0<c<1)$:%
\begin{equation}
Z_{g_{\text{\emph{sub}}}(t),\alpha }^{\ast }(X_{t}^{L})=e^{\mathcal{L}%
[R_{\alpha }(p)/c]}-1=:Z_{\text{\emph{sub}},\alpha }^{\ast }(X_{t}^{L}).
\label{Z_sub}
\end{equation}
\end{description}
\end{theorem}

See \cite{ADT2022CNS} for the details. In particular, 
\begin{equation}
Z_{\text{exp},1}^{\ast }(X_{t}^{L})=S_{BGS}(p)=H^{\ast }(X_{t}^{L})
\label{Remark82}
\end{equation}%
since $R_{1}(p)=S_{BGS}(p)$, Equation (\ref{S_BGS}), where we dispensed
again with factors. In other words, $Z_{g,\alpha }^{\ast }(X_{t}^{L})$
reduces to the conventional permutation entropy under the right assumptions.
This justifies calling $Z_{g,\alpha }^{\ast }(X_{t}^{L})$ permutation
entropy also in complexity classes other than the exponential class.

According to Theorem \ref{theo1}, $Z_{g,\alpha }^{\ast }(p)$ is composable,
i.e., there exists a formal group law $\Phi (x,y)$ such that%
\begin{equation}
\Phi (Z_{g,\alpha }^{\ast }(p),Z_{g,\alpha }^{\ast }(q))=Z_{g,\alpha }^{\ast
}(p\times q)).  \label{Compo}
\end{equation}%
Indeed, replace $\mathcal{W}(t)=e^{g(t)}$ and $\mathcal{W}%
^{-1}(s)=g^{-1}(\ln s)$ in Equation (\ref{phi_W}) to obtain%
\begin{equation}
\Phi (x,y)=g^{-1}[g(x+g^{-1}(0))+g(y+g^{-1}(0))]-g^{-1}(0).  \label{Phi*}
\end{equation}%
Plugging $g_{\text{\textrm{exp}}}(t)=ct$, $g_{\text{\textrm{fac}}}(t)=t\ln t$
and $g_{\text{\textrm{sub}}}(t)=ct\ln t$ into (\ref{Phi*}), it follows%
\begin{equation}
\Phi _{\text{\textrm{exp}}}(x,y)=x+y\text{,}  \label{Phi exp}
\end{equation}%
which amounts to the additivity of R\'{e}nyi's entropy, i.e.,%
\begin{equation}
R_{\alpha }(p\times q)=R_{\alpha }(p)+R_{\alpha }(q),  \label{R additivity}
\end{equation}%
and%
\begin{eqnarray}
\Phi _{\mathrm{fac}}(x,y) &=&\Phi _{\mathrm{sub}}(x,y)  \label{Phi fac} \\
&=&e^{\mathcal{L}[(x+1)\ln (x+1)+(y+1)\ln (y+1)]}-1.  \notag
\end{eqnarray}%
Use now the identity (\ref{identity}) and the additivity of $R_{\alpha }(p)$%
, Equation (\ref{R additivity}), to check the composability of $Z_{\mathrm{%
fac},\alpha }^{\ast }(p)$ and $Z_{\mathrm{sub},\alpha }^{\ast }(p)$. The
extensivity of $Z_{g,\alpha }^{\ast }(p)$ will be checked in the next
section.

\subsection{Permutation entropy rate}

\label{sec:42}

We know that $Z_{g,\alpha }^{\ast }(p)$ is extensive by construction
(Theorem \ref{theo1}). In this section we show that this property amounts to
the existence of the corresponding topological permutation entropy rate. On
the way we will find some interesting results.

Due to the Shannon-Khinchin Axioms SK2 (maximality) and SK3 (expansibility), 
$Z_{g,\alpha }^{\ast }(X_{t}^{L})$ (as any other generalized entropy for
that matter) achieves its maximum when the allowed $L$-patterns are
equiprobable, i.e., when $p=(p_{1},...,p_{L!})$ is of the form%
\begin{equation}
p_{i}=\left\{ 
\begin{array}{cl}
1/\mathcal{A}_{L}(\mathbf{X}) & \text{if the }i\text{th }L\text{-pattern is
allowed for }\mathbf{X}\text{ } \\ 
0 & \text{if the }i\text{th }L\text{-pattern is forbidden for }\mathbf{X}%
\end{array}%
\right.  \label{p_u}
\end{equation}%
for $i=1,...,L!$. In this particular case,%
\begin{equation}
R_{\alpha }(p)=\ln \mathcal{A}_{L}(\mathbf{X})  \label{R(p_u)}
\end{equation}%
for all $\alpha >0$. Plugging Equation (\ref{R(p_u)}) into (\ref{Zg_entropy}%
), we are led to the following definition.

\begin{definition}
The \emph{topological permutation entropy of order} $L$ of a process $%
\mathbf{X}$ of class $g$ is defined as 
\begin{equation}
Z_{g,0}^{\ast }(X_{t}^{L})\equiv Z_{g,0}^{\ast }(p_{u})=g^{-1}(\ln \mathcal{A%
}_{L}(\mathbf{X}))-g^{-1}(0),  \label{Z*_g,0}
\end{equation}%
where $p_{u}$ is the uniform probability distribution of allowed $L$%
-patterns for $\mathbf{X}$ as defined in Equation (\ref{p_u}).
\end{definition}

The notation $Z_{g,0}^{\ast }$ for the topological permutation entropy is
justified because $\ln \mathcal{A}_{L}(\mathbf{X})$ is formally obtained
from $R_{\alpha }(p)$, Equation (\ref{Renyi ent}), by setting $\alpha =0$.
Furthermore, with this notation $Z_{g,\alpha }^{\ast }$ denotes both the
topological version ($\alpha =0$) and the metric version ($\alpha >0$) of
the permutation entropy.

From%
\begin{equation}
R_{0}(p)\geq R_{\alpha }(p)  \label{R_0>R_alpha}
\end{equation}%
for all $\alpha >0$, it follows 
\begin{equation}
Z_{g,0}^{\ast }(p)\geq Z_{g,\alpha }^{\ast }(p)  \label{hierarch}
\end{equation}%
for all $\alpha >0$, since $g^{-1}(s)$ is a strictly increasing. From
definition (\ref{Z*_g,0}) and $\ln \mathcal{A}_{L}(\mathbf{X})\sim g(L)$,
Equation (\ref{pcf}), we obtain%
\begin{eqnarray}
\frac{Z_{g,\alpha }^{\ast }(p_{u})}{L} &=&\frac{Z_{g,0}^{\ast }(X_{t}^{L})}{L%
}=\frac{g^{-1}(\ln \mathcal{A}_{L}(\mathbf{X}))-g^{-1}(0)}{L}
\label{extensivityZ} \\
&\sim &\frac{g^{-1}(g(L))-g^{-1}(0)}{L}=\frac{L-g^{-1}(0)}{L}\sim 1_{.} 
\notag
\end{eqnarray}%
This shows that the \textquotedblleft extensivity
constant\textquotedblright\ of $Z_{g,\alpha }^{\ast }(p)$, namely $%
\lim_{L\rightarrow \infty }$ $Z_{g,0}^{\ast }(X_{t}^{L})/L$ (see Equation (%
\ref{ext})), is $1$.

\begin{remark}
\label{Differentiability_g}In\ the definition (\ref{Zg_entropy}) of $%
Z_{g,\alpha }^{\ast }(p)$ we dispensed with the factor $\lambda $ from
Equation (\ref{maineq}). Had we included that factor in the definition of $%
Z_{g,\alpha }^{\ast }(p)$, the extensivity constant would be precisely $%
\lambda =g^{\prime }(g^{-1}(0))$, see Equation (\ref{lambda}). In that case,
both if $g(t)=ct$ (exponential class for $c>0$) and $g(t)=ct\ln t$
(factorial class for $c=1$ and sub-factorial class for $0<c<1$), the
extensivity constant is $g^{\prime }(g^{-1}(0))=c$.
\end{remark}

To get rid of the dependence of $Z_{g,\alpha }^{\ast }(X_{t}^{L})$ on $L$,
we turn to the entropy rate per variable, $Z_{g,\alpha }^{\ast
}(X_{t}^{L})/L $, and take the limit when $L\rightarrow \infty $.

\begin{definition}
The \emph{permutation entropy rate}\textit{\ }(or just \emph{permutation
entropy})\textit{\ of a process} $\mathbf{X}$ of class $g$ is defined as 
\begin{equation}
z_{g,\alpha }^{\ast }(\mathbf{X})=\lim_{L\rightarrow \infty }\frac{1}{L}%
Z_{g,\alpha }^{\ast }(X_{t}^{L}),  \label{Z*}
\end{equation}%
where $\alpha \geq 0$. For $\alpha >0$, $z_{g,\alpha }^{\ast }(\mathbf{X})$
is the $\emph{metric}$ permutation entropy of $\mathbf{X}$, while $%
z_{g,0}^{\ast }(\mathbf{X})$ is the \emph{topological} permutation entropy
of $\boldsymbol{X}$.
\end{definition}

As a matter of fact, the existence of $z_{g,\alpha }^{\ast }(\mathbf{X})$
amounts to the extensivity of $Z_{g,\alpha }^{\ast }(X_{t}^{L})$. From (\ref%
{hierarch}) and (\ref{extensivityZ}) it follows 
\begin{equation}
z_{g,\alpha }^{\ast }(\mathbf{X})\leq z_{g,0}^{\ast }(\mathbf{X})=1
\label{bounds2}
\end{equation}%
for every complexity class $g$ and $\alpha >0$. In other words, the values
of the entropy rate $z_{g,\alpha }^{\ast }(\mathbf{X})$, $\alpha >0$, are
restricted to the interval $[0,1]$.

\begin{remark}
\label{RemarkLast}\label{Remark WN}For white noise (WN), the probability
distribution of the $L$-patterns is uniform for every $L$. Therefore, 
\begin{equation}
z_{\text{\emph{fac}},\alpha }^{\ast }(\text{WN})=z_{\text{\emph{fac}}%
,0}^{\ast }(\text{WN})=1  \label{Remark15c}
\end{equation}%
for all $\alpha >0$ by Equation (\ref{bounds2}).
\end{remark}

\begin{figure*}[tbp]
\begin{center}
\includegraphics[scale=0.43]{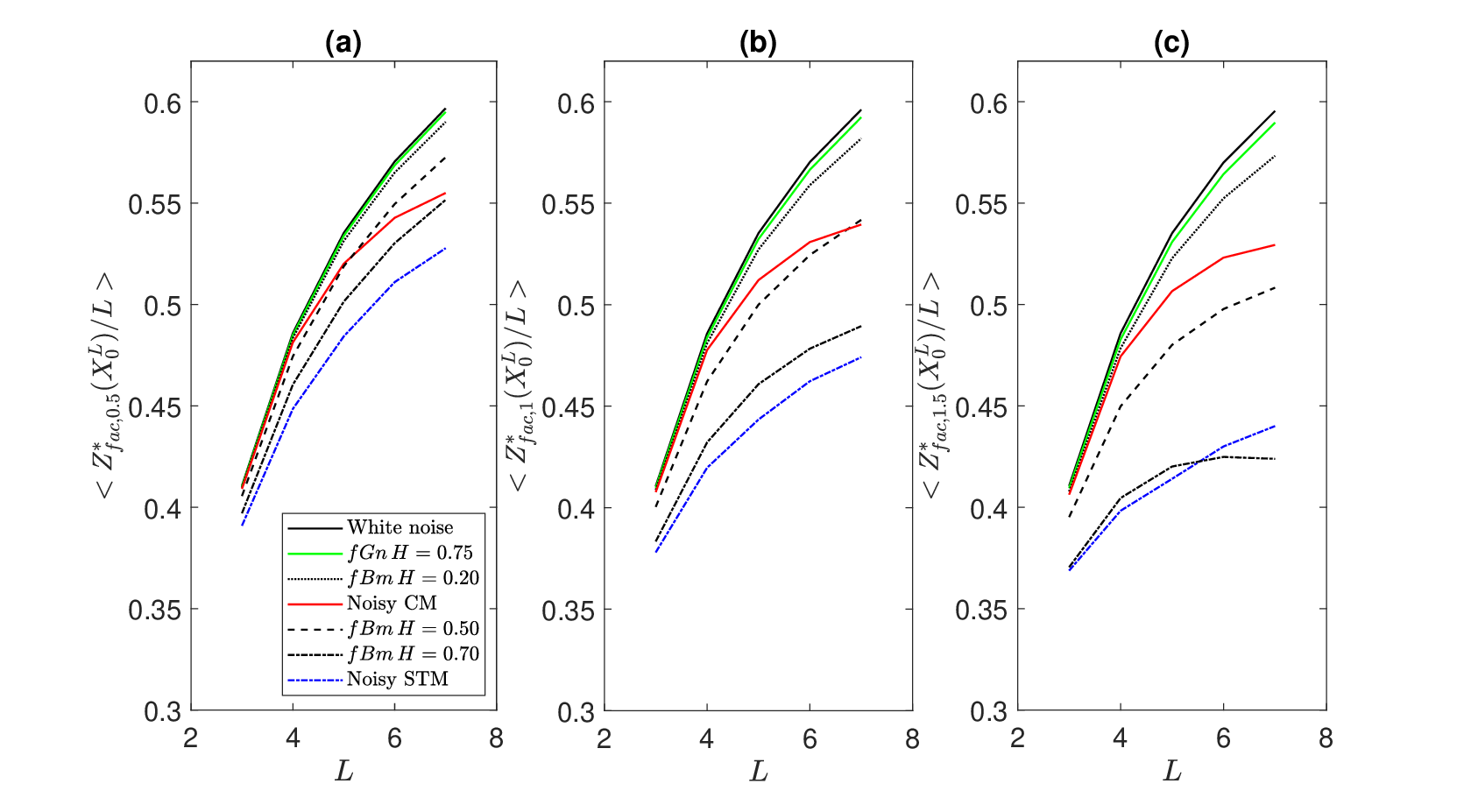}
\end{center}
\caption{Averages of $Z_{\text{fac},0.5}^{\ast }(X_{0}^{L})/L$ (a), $Z_{%
\text{fac},1}^{\ast }(X_{0}^{L})/L$ (b) and $Z_{\text{fac},1.5}^{\ast
}(X_{0}^{L})/L$ (c) over 10 realizations of the random processes listed in
the inset (see Section \protect\ref{sec:33} for detail) are plotted vs $L$
for $3\leq L\leq 7$.}
\label{fig:Figura4}
\end{figure*}

Finally, Figure~\ref{fig:Figura4} shows the entropies $Z_{\text{fac},\alpha
}^{\ast }(X_{0}^{L})$ per variable for $3\leq L\leq 7$ and $\alpha
=0.5,1,1.5 $. The most interesting features of this figure can be summarized
as follows: (i) The curve for white noise is the maximal envelope of $Z_{%
\text{fac},\alpha }^{\ast }(X_{0}^{L})$/$L=(\exp \mathcal{L}[R_{\alpha
}(X_{0}^{L})]-1)/L$ (see (\ref{Z_fac})) because then $R_{\alpha
}(X_{0}^{L})=\ln L!=R_{0}(X_{0}^{L})$ for $\alpha >0$, which is the maximum
of $R_{\alpha }(X_{0}^{L})$ for each $L$; (ii) As $L$ increases, longer
ranges of dependencies between variables can be captured by $Z_{\text{fac}%
,\alpha }^{\ast }(X_{t}^{L})$, which explains why curves of different
processes can cross; (iii) The curves are further separated the greater $%
\alpha $ is. The last comment shows that the parameter $\alpha $ is useful
when it comes to the characterization and discrimination of processes.

\section{Conclusion}

\label{sec:5}

In this paper we have revisited the concepts and main properties of group
entropy (Section \ref{sec:2}) and permutation complexity (Section \ref{sec:3}%
), as well as some applications of group entropies to permutation
complexity, the most important being the extension of permutation entropy
from deterministic processes to random processes. In a nutshell, a group
entropy is a generalized entropy (i.e., a positive functional on probability
distributions that satisfies the first three Shannon-Khinchin axioms) that,
additionally, is composable in the sense of Definition \ref{composab}.
Permutation entropy is the Shannon entropy of the probability distribution
of ordinal patterns (the main character of permutation complexity) obtained
from a deterministic or random real-valued process, but its rate diverges
for random process. The explanation for that is simple:\ the number of
allowed ordinal patterns grows exponentially with their length for
deterministic processes (as happens in information theory with the number of
words and their length), while it does so super-exponentially (in fact,
factorially) for the usual random processes. It is precisely at this point
where group entropy is called for.

Indeed, certain group entropies, called $Z$-entropies, are so defined that
they are extensive, see Equation (\ref{ext}), for all systems belonging to
any given complexity class. These complexity classes are defined by the
state space growth rate function $\mathcal{W}(N)$, see Section \ref{sec:24}.
The extensivity of the $Z$-entropy of a complexity class amounts to the
convergence of its rate for all systems in that class. The translation of
this general setting to permutation complexity (mostly related to nonlinear
time series analysis) is smooth. Thus, the $Z$-entropy for the exponential
complexity class (i.e., systems with a exponential growth of $\mathcal{W}(N)$%
) coincides with the conventional permutation R\'{e}nyi entropy $R_{\alpha
}(p)$, Equation (\ref{Renyi ent}), which includes the conventional
permutation (Shannon) entropy for $\alpha =1$. And the extensivity of the $Z$%
-entropy for the factorial complexity class (i.e., systems with a factorial
growth of $\mathcal{W}(N)$) translates into the convergence of the
corresponding, complexity-based permutation entropy rate for random
processes. The result is a new conceptual route to complexity in time series
analysis.

\section{Acknowledgements}

We are very grateful to our Reviewers for their comments and suggestions,
which have helped to improve the original version of our manuscript. J.M.A.
and R.D. were financially supported by the Spanish grant
PID2019-108654GB-I00/AEI/10.13039/501100011033. J.M.A. was also supported by
Generalitat Valenciana, Spain, grant PROMETEO/2021/063. The research of P.T.
has been supported by the research project PGC2018-094898-B-I00, Ministerio
de Ciencia, Innovaci\'{o}n y Universidades and Agencia Estatal de Investigaci%
\'{o}n, Spain, and by the Severo Ochoa Programme for Centres of Excellence
in R\&D (CEX2019-000904-S), Ministerio de Ciencia, Innovaci\'{o}n y
Universidades y Agencia Estatal de Investigaci\'{o}n, Spain. P.T. is member
of the Gruppo Nazionale di Fisica Matematica (GNFM).

\bigskip\ 

\end{document}